\documentclass[prd,preprintnumbers,amsmath,amssymb]{revtex4}
\usepackage{cancel}
\usepackage{booktabs}
\usepackage{multirow}
\usepackage{slashed}
\usepackage{graphicx}
\usepackage{appendix}

\usepackage[colorlinks,
            citecolor=blue,
            anchorcolor=red,
            menucolor=red,
            linkcolor=red,
            filecolor=red,
            runcolor=red,
            urlcolor=blue,
            frenchlinks=red]{hyperref}

\begin{document}
\title{Systematic studies of charmonium-, bottomonium-, and $B_c$-like tetraquark states }
\author{Jing Wu$^1$} \email{wujing18@sdjzu.edu.cn}\affiliation{$^1$School of Science, Shandong Jianzhu University, Jinan 250101, China}
\author{Xiang Liu$^{2,3}$}
\email{xiangliu@lzu.edu.cn} \affiliation{
$^2$School of Physical Science and Technology, Lanzhou University, Lanzhou 730000, China\\
$^3$Research Center for Hadron and CSR Physics, Lanzhou University
and Institute of Modern Physics of CAS, Lanzhou 730000, China }
\author{Yan-Rui Liu$^4$}\email{ yrliu@sdu.edu.cn} \affiliation{ $^4$School of Physics, Shandong University, Jinan 250100, China}
\author{Shi-Lin Zhu$^{5,6,7}$}
\email{zhusl@pku.edu.cn} \affiliation{
$^5$School of Physics and State Key Laboratory of Nuclear Physics and Technology, Peking University, Beijing 100871, China\\
$^6$Collaborative Innovation Center of Quantum Matter, Beijing 100871, China\\
$^7$Center of High Energy Physics, Peking University, Beijing
100871, China }

\begin{abstract}
We study the mass splittings of $Q_1q_2\bar{Q}_3\bar{q}_4$ ($Q=c,b$,
$q=u,d,s$) tetraquark states with chromomagnetic interactions
between their quark components. Assuming that $X(4140)$ is the
lowest $J^{PC}=1^{++}$ $cs\bar{c}\bar{s}$ tetraquark, we estimate
the masses of the other tetraquark states. From the obtained masses
and defined measure reflecting effective quark interactions, we find
the following assignments for several exotic states: (1) both
$X(3860)$ and the newly observed $Z_c(4100)$ seem to be $0^{++}$
$cn\bar{c}\bar{n}$ tetraquarks; (2) $Z_c(4200)$ is probably a
$1^{+-}$ $cn\bar{c}\bar{n}$ tetraquark; (3) $Z_c(3900)$, $X(3940)$,
and $X(4160)$ are unlikely compact tetraquarks; (4) $Z_c(4020)$ is
unlikely a compact tetraquark, but seems the hidden-charm
correspondence of $Z_b(10650)$ with $J^{PC}=1^{+-}$; and (5)
$Z_c(4250)$ can be a tetraquark candidate but the quantum numbers
cannot be assigned at present. We hope further studies may check the
predictions and assignments given here.
\end{abstract}

\date{\today}

\maketitle
\section{Introduction}\label{sec1}

A hot topic in hadron physics study is to identify multiquark states
from the observed exotic structures. Through explorations on their
masses, productions, and decay properties, we may understand the
problem how the strong interaction forces nonobservable quarks and
gluons to form observable hadrons. Before 2003, the situation in
understanding hadron structures was simple because the quark model
gave a successful and satisfactory description for hadron spectra
\cite{Godfrey:1985xj}, although there exist a few hadrons difficult
to understand. In 2003, experimentalists opened the Pandora's box
for exotic states through the observation of $X(3872)$
\cite{Choi:2003ue}. Since then, more and more unexpected XYZ states
were observed and the situation for hadron physics study became
complicated
\cite{Liu:2013waa,Olsen:2014qna,Chen:2016qju,Richard:2016eis,Esposito:2016noz,Lebed:2016hpi,Guo:2017jvc,Ali:2017jda,Olsen:2017bmm,Karliner:2017qhf}.
To understand a little more the above mentioned problem, the
discussions in this work aim at basic features of ground
charmonium-like, bottomonium-like, and $B_c$-like tetraquark states
with even $P$-parities.

As the first exotic charmonium-like state above the $D\bar{D}$
threshold, the $X(3872)$ motivated heated discussions on its nature
\cite{Chen:2016qju,Ali:2017jda}. Its $J^{PC}$ are determined to be
$1^{++}$ but the mass is tens of MeV lower than the quark model
prediction if it is a charmonium. Since the meson is extremely close
to the $D^0\bar{D}^{*0}$ threshold, it is widely regarded as a
loosely bound $D\bar{D}^*$ molecule. Discussions in the tetraquark
picture and hybrid picture are also performed. However, without a
$c\bar{c}$ core, it is difficult to understand the measured ratios
${\cal B}(X(3872)\to\Psi(2S)\gamma):{\cal B}(X(3872)\to
J/\Psi\gamma)=2.46\pm0.64\pm0.29$ by LHCb \cite{Aaij:2014ala}. The
$X(3872)$ seems to be a charmonium affected significantly by the
$D\bar{D}^*$ threshold \cite{Swanson:2004pp,Li:2009zu}. Anyway, one
cannot identify this exotic meson as a pure tetraquark state.

To identify multiquark states, we may look for structures according
to several ideas. The easiest approach is to study structures with
explicitly exotic quantum numbers, e.g. charged charmonium-like or
bottomonium-like states. The quark content of the charged
hidden-charm or hidden-bottom mesons should be at least four if
their nonresonance interpretations are excluded. Up to now,
experiments have observed several charged structures, $Z_c(4430)$
\cite{Choi:2007wga,Chilikin:2013tch,Aaij:2014jqa}, $Z_c(4050)$
\cite{Mizuk:2008me}, $Z_c(4250)$ \cite{Mizuk:2008me}, $Z_c(3900)$
\cite{Ablikim:2013mio,Liu:2013dau,Xiao:2013iha,Ablikim:2015tbp,Collaboration:2017njt},
$Z_c(3885)$ \cite{Ablikim:2013xfr,Ablikim:2015gda,Ablikim:2015swa},
$Z_c(4020)$ \cite{Ablikim:2013wzq,Ablikim:2014dxl}, $Z_c(4025)$
\cite{Ablikim:2013emm,Ablikim:2015vvn}, and so on. Very recently,
LHCb found the evidence for a charged charmonium-like resonance
$Z_c^-(4100)$ in the decay $B^0\to Z_c^-(4100)K^+\to\eta_c\pi^-K^+$
\cite{Aaij:2018bla}. The measured mass and width are
$4096\pm20^{+18}_{-22}$ MeV and $152\pm 58^{+60}_{-35}$ MeV,
respectively. Its possible quantum numbers are $J^P=0^+$ or $1^-$.
They are certainly four-quark state candidates. However, it is not
easy to justify whether they are compact tetraquarks or meson-meson
molecules. In this paper, we will try to understand whether parts of
these charged states are compact tetraquarks or just molecules.

It is also possible to identify a multiquark state from its high
mass that a conventional hadron cannot have. The observed
$P_c(4380)$ and $P_c(4450)$ by the LHC Collaboration
\cite{Aaij:2015tga} are two such states. They look like excited
nucleons but can be identified as pentaquark states because an
orbital or radial excitation energy larger than 3 GeV for light quarks is an unnatural interpretation for the high masses while the creation of a $c\bar{c}$ pair can naturally explain. Ref. \cite{Wu:2010jy} predicted the existence of hidden-charm
pentaquarks with this idea. Similarly, one may identify other high
mass states looking like conventional hadrons as multiquark states
if experiments could observe them. However, one still cannot easily
distinguish compact tetraquarks from molecules except the
$QQ\bar{Q}\bar{q}$ case \cite{Cui:2006mp,Chen:2016ont,Jiang:2017tdc}
in this possibility.

If experiments could observe an exotic structure that the molecule
picture is not applicable, it is possible to identify it as a
compact tetraquark. In Refs. \cite{D0:2016mwd,Abazov:2017poh}, the
D0 Collaboration claimed an exotic $B_s^0\pi^\pm$ state and named it
$X(5568)$. This meson contains four different flavors. From its low
mass ($\sim$200 MeV lower than the $B\bar{K}$ threshold), the
$X(5568)$ is unlikely a molecule. If it really exists, it might be a
compact tetraquark. Unfortunately, the LHCb \cite{Aaij:2016iev}, CMS
\cite{Sirunyan:2017ofq}, CDF \cite{Aaltonen:2017voc}, and ATLAS
\cite{Aaboud:2018hgx} Collaborations did not confirm this state. The
identification of compact tetraquarks along this idea has not been
achieved yet.

We have one more possibility to identify compact multiquarks through
number of states. The exotic structure $X(4140)$ was first observed
by the CDF Collaboration \cite{Aaltonen:2009tz} in the invariant
mass distribution of $J/\psi \phi$. In the latter measurements with
the same channel by various collaborations
\cite{Chatrchyan:2013dma,Abazov:2013xda,Abazov:2015sxa,Aaij:2016nsc,Aaij:2016iza},
LHCb confirmed the $X(4140)$, determined its quantum numbers to be
$J^{PC}=1^{++}$, established another $1^{++}$ state $X(4274)$, and
observed two more $0^{++}$ structures $X(4500)$ and $X(4700)$. The
existence of two $1^{++}$ states does not support the molecule
interpretations for them \cite{Aaij:2016iza}. On the other hand, the
$cs\bar{c}\bar{s}$ tetraquark configuration can account for such an
observation \cite{Stancu:2009ka}. This picture also favors the
assignment for the Belle $X(4350)$ \cite{Shen:2009vs} as their
$0^{++}$ tetraquark partner \cite{Wu:2016gas}. In this paper, we
identify the $X(4140)$ as the lowest $1^{++}$ $cs\bar{c}\bar{s}$
tetraquark state and use its mass as an input to estimate the masses
of other charmonium-, bottomonium-, and $B_c$-like tetraquark states.

This paper is organized as follows. After the introductory Sec.
\ref{sec1}, we present the theoretical formalism in Sec. \ref{sec2}
by showing necessary wave functions and Hamiltonian matrices. In Sec. \ref{sec3}, we determine model parameters,
present strategy for the estimation of tetraquark masses, list numerical results, analyze possible assignments for the
observed exotic mesons, and predict possible tetraquarks. The last section is for discussions and summary.

\section{Formalism}\label{sec2}

In this article, we use the notation $Q_1q_2\bar{Q}_3\bar{q}_4$
($Q=c,b$; $q=n,s$; $n=u,d$) to generally denote the considered
system. If the system is truly neutral, $Q_1=Q_3=Q,\,q_2=q_4=q$ and
the notation becomes $Qq\bar{Q}\bar{q}$. From the $SU(3)_f$
symmetry, the tetraquarks belong to $8_f$ and $1_f$ representations.
Since the flavor symmetry is broken, the isoscalar states would mix
with some angle. In principle, the resulting flavor wave functions
of the physical $I=0$ states contain both $Q_1n\bar{Q}_3\bar{n}$ and
$Q_1s\bar{Q}_3\bar{s}$ parts. At present, we just consider the ideal
mixing case, i.e. $Q_1n\bar{Q}_3\bar{n}$ and $Q_1s\bar{Q}_3\bar{s}$
do not mix.

The effective Hamiltonian in the adopted chromomagnetic interaction (CMI) model reads,
\begin{equation}\label{equation:CMI}
H=\sum_{i}m_i+{H}_{CM}=\sum_{i}m_i-\sum_{i<j}C_{ij}\widetilde{\lambda}_i\cdot\widetilde{\lambda}_j\sigma_i\cdot\sigma_j,
\end{equation}
where $\widetilde{\lambda}_i=\lambda_i$ $(-\lambda^*_i)$ for quarks (antiquarks). The involved parameters are only effective coupling constants $C_{ij}$ and effective masses $m_i$ containing various effects. This Hamiltonian is reduced from a realistic model, which can be found in Refs. \cite{DeRujula:1975qlm,Zhou:2018pcv}. Then the formula for the mass estimation is
\begin{eqnarray}\label{mass}
M=\sum_{i=1} m_i+\langle {H}_{CM}\rangle.
\end{eqnarray}
In calculating the last term, we use the diquark-antidiquark bases to express the wave functions for the S-wave $Q_1q_2\bar{Q}_3\bar{q}_4$ systems whose $P$-parities are always positive. Here, the notation ``diquark'' just means two quarks and does not mean a
compact substructure. If one uses the meson-meson bases, the same eigenvalues after
diagonalization will be obtained. In the present case, the Pauli principle has no restriction
on the total wave functions, but one should notice the possible $C$-parity once a state is truly neutral. The involved color (spin) wave functions $\phi_{1,2}$ ($\chi_{1,2,\cdots,6}$) are
\begin{eqnarray}
\phi_1&=&|\bar{3}_c,3_c,1_c\rangle,\quad \phi_2=|6_c,\bar{6}_c,1_c\rangle,\nonumber\\
\chi_1&=&|1_S,1_S,2_S\rangle,\quad \chi_2=|1_S,1_S,1_S\rangle,\quad \chi_3=|1_S,1_S,0_S\rangle,\nonumber\\
\chi_4&=&|1_S,0_S,1_S\rangle,\quad \chi_5=|0_S,1_S,1_S\rangle,\quad \chi_6=|0_S,0_S,0_S\rangle,
\end{eqnarray}
where the color representations (spins) in order in $\phi_i$ ($\chi_j$) are for diquark, antidiquark, and system, respectively.
We define the total wave function as
\begin{eqnarray}
\phi_i\chi_j\equiv (Q_1q_2\bar{Q}_3\bar{q}_4)\otimes\phi_i\otimes\chi_j.
\end{eqnarray}

Compared with the $cs\bar{c}\bar{s}$ case where a $C$-parity can be given, the CMI matrices in the present cases are the generalized ones in Ref. \cite{Wu:2016gas}. Now we have
\begin{eqnarray}\label{matrixJ2}
\langle H_{CM}\rangle_{J^P=2^+}=\left(
\begin{array}{cc}
\frac{4}{3}(2\tau+\alpha) & 2\sqrt{2}\nu \\
& \frac{2}{3}(5\alpha-2\tau)
\end{array}\right),
\end{eqnarray}
\begin{eqnarray}\label{matrixJ0}
\langle H_{CM}\rangle_{J^P=0^+}=\left(\begin{array}{cccc}
 \frac{8}{3}(\tau-\alpha) & 4\sqrt{2} \nu &-\frac{4}{\sqrt{3}}\nu & 2 \sqrt{6}\alpha\\
  & -\frac{4}{3}(\tau+5\alpha) & 2\sqrt{6}\alpha & -\frac{10}{\sqrt{3}}\nu\\
& & -8 \tau & 0 \\
  & &  & 4 \tau
\end{array}\right),
\end{eqnarray}
and
\begin{eqnarray}\label{matrixJ1}
\langle H_{CM}\rangle_{J^P=1^+}=\left(\begin{array}{cccccc} \frac{4}{3} (2\tau-\alpha) & \frac{4\sqrt{2}}{3}\beta & -\frac{4\sqrt{2}}{3} \mu & 2\sqrt{2} \nu & -4\mu & 4 \beta \\
& \frac{8}{3} (2\theta-\tau) & \frac43\nu & -4\mu & 0 & -2\sqrt{2}\alpha \\
& & -\frac83(\tau+2\theta) & 4\beta & -2\sqrt{2}\alpha & 0 \\
& & & -\frac23(2\tau+5\alpha) &\frac{10\sqrt{2}}{3}\beta & -\frac{10\sqrt{2}}{3}\mu \\
& & & & \frac43(\tau-2\theta) & \frac{10}{3} \nu \\
& & & & & \frac43(\tau+2\theta)
\end{array}\right),
\end{eqnarray}
where the defined variables are
\begin{eqnarray}\label{definedvariables}
\tau&=&C_{12}+C_{34},\quad \theta=C_{12}-C_{34},\nonumber\\
\alpha&=&C_{13}+C_{24}+C_{14}+C_{23},\nonumber\\
\beta&=&C_{13}-C_{24}-C_{14}+C_{23},\nonumber\\
\mu&=&C_{13}-C_{24}+C_{14}-C_{23},\nonumber\\
\nu&=&C_{13}+C_{24}-C_{14}-C_{23}
\end{eqnarray}
and the corresponding base vectors for the matrices are $(\phi_1\chi_1,\phi_2\chi_2)^T$, $(\phi_1\chi_3,~\phi_2\chi_3,~\phi_1\chi_6,~\phi_2\chi_6)^T$, and $(\phi_1\chi_2,\phi_1\chi_4,\phi_1\chi_5,\phi_2\chi_2,\phi_2\chi_4,\phi_2\chi_5)^T$, respectively. When the considered state is truly neutral, the matrices for the cases $J^{PC}=2^{++}$ and $0^{++}$ are the same as above, but that for the
case $J^{PC}=1^{++}$ is
\begin{eqnarray}\label{matrixJ1pp}
\langle H_{CM}\rangle=\left(
\begin{array}{cc}
-\frac{4}{3}(4C_{Qq}-C_{Q\bar{Q}}-C_{q\bar{q}}+2C_{Q\bar{q}})& -2\sqrt{2}(C_{Q\bar{Q}}+C_{q\bar{q}}+2C_{Q\bar{q}}) \\
  & \frac{2}{3}(4C_{Qq}+5C_{Q\bar{Q}}+5C_{q\bar{q}}-10C_{Q\bar{q}})
\end{array}\right)
\end{eqnarray}
and that for the case $J^{PC}=1^{+-}$ is
\begin{eqnarray}\label{matrixJ1pm}
\langle H_{CM}\rangle=\left(\begin{array}{cccc}
\frac43\left(\begin{array}{c}4C_{Qq}-2C_{Q\bar{q}}\\-C_{Q\bar{Q}}-C_{q\bar{q}}\end{array}\right)& 2\sqrt2\left(\begin{array}{c}C_{Q\bar{Q}}+C_{q\bar{q}}\\-2C_{Q\bar{q}}\end{array}\right) & \frac83 (C_{Q\bar{Q}}-C_{q\bar{q}}) & -4\sqrt2(C_{Q\bar{Q}}-C_{q\bar{q}})\\
 & -\frac23\left(\begin{array}{c}4C_{Qq}+10C_{Q\bar{q}}\\+5C_{Q\bar{Q}}+5C_{q\bar{q}}\end{array}\right) & -4\sqrt2(C_{Q\bar{Q}}-C_{q\bar{q}}) & \frac{20}{3}(C_{Q\bar{Q}}-C_{q\bar{q}})\\
  & & -\frac43\left(\begin{array}{c}4C_{Qq}-2C_{Q\bar{q}}\\+C_{Q\bar{Q}}+C_{q\bar{q}}\end{array}\right) & 2\sqrt2\left(\begin{array}{c}C_{Q\bar{Q}}+C_{q\bar{q}}\\+2C_{Q\bar{q}}\end{array}\right) \\
  & & & \frac23\left(\begin{array}{c}4C_{Qq}+10C_{Q\bar{q}}\\-5C_{Q\bar{Q}}-5C_{q\bar{q}}\end{array}\right)
\end{array}\right).
\end{eqnarray}
Their corresponding base vectors are $(\phi_1\chi_e,~\phi_2\chi_e)^T$ and $(\phi_1\chi_2,~\phi_2\chi_2,~\phi_1\chi_o,~\phi_2\chi_o)^T$, respectively. Here $\phi_i\chi_e$ ($\phi_i\chi_o$) represents $C$-even ($C$-odd) wave function. The forms of such wave functions are similar to those obtained in Ref. \cite{Wu:2016gas}. Since we also consider the color structure $|6_c,\bar{6}_c,1_c\rangle$ for the tetraquarks, the above Eqs. \eqref{matrixJ2} to \eqref{matrixJ1pm} can be actually thought of as generalizations of those for $|\bar{3}_c,3_c,1_c\rangle$ tetraquarks given in Ref. \cite{Maiani:2004vq}.

\section{Numerical analysis}\label{sec3}

\subsection{Parameters and estimation strategy}

The parameters in the CMI model are effective masses of the quarks
and coupling strengths between quark components. We need 14 coupling
strengths in the present study: $C_{cn}$, $C_{cs}$, $C_{bn}$,
$C_{bs}$, $C_{c\bar{n}}$, $C_{c\bar{s}}$, $C_{b\bar{n}}$,
$C_{b\bar{s}}$, $C_{c\bar{c}}$, $C_{b\bar{b}}$, $C_{c\bar{b}}$,
$C_{n\bar{n}}$, $C_{s\bar{s}}$, and $C_{n\bar{s}}$. Most of them can be extracted from the measured masses \cite{Tanabashi:2018oca} of the low-lying conventional hadrons (see table \ref{parameter}), but the determination of $C_{s\bar{s}}$ and $C_{c\bar{b}}$ needs
approximations. We here assume $C_{s\bar{s}}=C_{ss}C_{n\bar{n}}/C_{nn}=10.5$ MeV and adopt
$C_{c\bar{b}}=3.3$ MeV extracted from $M_{B_c^*}-M_{B_c}=70$ MeV
\cite{Godfrey:1985xj}. Parts of spectroscopic coupling parameters have been derived in Ref. \cite{Maiani:2004vq}. The values of our coupling parameters are consistent with those in that paper, see discussions in Refs. \cite{Wu:2016gas,Li:2018vhp}. The effective quark masses we extracted are $m_n=361.7$ MeV, $m_s=540.3$ MeV, $m_c=1724.6$ MeV, and $m_b=5052.8$
MeV, which are close to those obtained in Ref. \cite{Karliner:2014gca}.

\begin{table}[htbp]
\caption{Chromomagnetic interactions for various hadrons and
obtained effective coupling constants in units of MeV.
}\label{parameter} \centering
\begin{tabular}{cccccc}\hline\hline
Hadron&$\langle H_{CM}\rangle$&Hadron&$\langle
H_{CM}\rangle$&$C_{ij}$\\\hline
$N$&$-8C_{nn}$&$\Delta$&$8C_{nn}$&$C_{nn}=18.4$\\
$\Sigma$&$\frac{8}{3}C_{nn}-\frac{32}{3}C_{n s}$&$\Sigma^*$&$\frac{8}{3}C_{nn}+\frac{16}{3}C_{n s}$&$C_{n s}=12.4$\\
$\Xi^0$&$\frac{8}{3}(C_{ss}-4C_{n s})$&$\Xi^{*0}$&$\frac{8}{3}(C_{ss}+C_{n s})$&\\
$\Omega$&8$C_{ss}$&&&$C_{ss}=6.5$\\
$\Lambda$&$-8C_{nn}$\\
$\pi$&$-16C_{n\bar{n}}$&$\rho$&$\frac{16}{3}C_{n\bar{n}}$&$C_{n\bar{n}}=29.8$\\
$K$&$-16C_{n\bar{s}}$&$K^*$&$\frac{16}{3}C_{n\bar{s}}$&$C_{n\bar{s}}=18.7$\\
$D$&$-16C_{c\bar{n}}$&$D^{*}$&$\frac{16}{3}C_{c\bar{n}}$&$C_{c\bar{n}}=6.7$\\
$D_s$&$-16C_{c\bar{s}}$&$D_{s}^{*}$&$\frac{16}{3}C_{c\bar{s}}$&$C_{c\bar{s}}=6.7$\\
$B$&$-16C_{b\bar{n}}$&$B^{*}$&$\frac{16}{3}C_{b\bar{n}}$&$C_{b\bar{n}}=2.1$\\
$B_s$&$-16C_{b\bar{s}}$&$B^{*}$&$\frac{16}{3}C_{b\bar{s}}$&$C_{b\bar{s}}=2.3$\\
$\eta_{c}$&$-16C_{c\bar{c}}$&$J/\psi$&$\frac{16}{3}C_{c\bar{c}}$&$C_{c\bar{c}}=5.3$\\
$\eta_{b}$&$-16C_{b\bar{b}}$&$\Upsilon$&$\frac{16}{3}C_{b\bar{b}}$&$C_{b\bar{b}}=2.9$\\
$\Sigma_{c}$&$\frac{8}{3}C_{nn}-\frac{32}{3}C_{cn}$&$\Sigma_{c}^*$&$\frac{8}{3}C_{nn}+\frac{16}{3}C_{cn}$&$C_{cn}=4.0$\\
$\Xi'_{c}$&$\frac{8}{3}C_{n s}-\frac{16}{3}C_{cn}-\frac{16}{3}C_{cs}$&$\Xi_{c}^*$&$\frac{8}{3}C_{n s}+\frac{8}{3}C_{cn}+\frac{8}{3}C_{cs}$&$C_{cs}=4.5$\\
$\Sigma_{b}$&$\frac{8}{3}C_{nn}-\frac{32}{3}C_{bn}$&$\Sigma_{b}^{*}$&$\frac{8}{3}C_{nn}+\frac{16}{3}C_{bn}$&$C_{bn}=1.3$\\
$\Xi'_{b}$&$\frac{8}{3}C_{n
s}-\frac{16}{3}C_{bn}-\frac{16}{3}C_{bs}$&$\Xi_b^*$&
$\frac{8}{3}C_{n s}+\frac{8}{3}C_{bn}+\frac{8}{3}C_{bs}$&$C_{bs}=1.2$\\
\hline
\end{tabular}
\end{table}

When one substitutes these parameters into the mass formula \eqref{mass}, the tetraquark masses may be estimated. However, if we check the
numerical values for the masses of the conventional hadrons with
this formula and the above parameters, deviations from experimental
results are found (see table IV of Ref. \cite{Zhou:2018pcv}).
Usually, the obtained masses are larger than the measured values,
which indicates that the attractions between quark components are
not sufficiently considered in the simple model. The application of
this formula to multiquark states should also lead to higher masses than those they should be.
On the theoretical side, such values can be treated as upper limits
of the tetraquark masses.

The reason for the overestimated masses is because of the adopted assumption that the above extracted parameters are applicable to every system. In principle, each system has its own values of parameters. From the reduction procedure for the model Hamiltonian and the fact that the spacial wave functions are not the same for different systems, this assumption certainly induces uncertainties. The uncertainties in coupling strengths affect the mass splittings between the considered tetraquark states and the effects should not be large. On the other hand, the uncertainties in the effective quark masses affect the mass shifts of the states, which may be significant. 
To reduce the uncertainties in mass estimation, we adopt
another method by introducing a reference system and modifying the
mass formula to be
\begin{eqnarray}\label{massref}
 M=(M_{ref}-\langle{H}_{CM}\rangle_{ref})+\langle{H}_{CM}\rangle.
\end{eqnarray}
Here, $M_{ref}$ and $\langle{H}_{CM}\rangle_{ref}$ are the
physical mass of the reference system and the corresponding CMI
eigenvalue, respectively. For $M_{ref}$, one may use the mass of a
reference multiquark state or use the threshold of a reference
hadron-hadron system whose quark content is the same as the
considered multiquark states. With this method, the problem of using
extracted quark masses from conventional hadrons in multiquark
systems \cite{Stancu:2009ka} is evaded and part of missed
attractions between quark components is phenomenologically
compensated. In previous studies
\cite{Chen:2016ont,Wu:2016gas,Zhou:2018pcv,Wu:2016vtq,Li:2018vhp,Hyodo:2012pm,Wu:2017weo,Luo:2017eub},
we mainly adopted hadron-hadron thresholds. One finds that the
estimated multiquark masses with this method are always lower than
those with Eq. \eqref{mass}. Since the number of thresholds may be
more than 1, there is a question which threshold leads to more
reasonable masses. As a multibody system, the size of a tetraquark
state should be larger than that of a conventional hadron and the
distance between two quark components in tetraquarks may be larger
than that in a conventional meson. The resulting effect is that the
attraction between quark components should be weaker. Thus, although
we cannot give a definite answer, probably the meson-meson threshold
leading to higher masses gives more reasonable tetraquark masses. In
the present study, besides the possible hadron-hadron thresholds, we
may additionally turn to $X(4140)$ by assuming it as the ground
$cs\bar{c}\bar{s}$ tetraquark state with $J^{PC}=1^{++}$. It seems
that using $X(4140)$ as an input is a better approach than the
adoption of meson-meson thresholds. In Ref. \cite{Wu:2016gas}, we
have performed the exploration for the $cs\bar{c}\bar{s}$ states
with this input and gotten higher masses than with the
$D_s\bar{D}_s$ threshold. This observation probably indicates that the highest
masses estimated with various hadron-hadron thresholds are still
lower than the tetraquark masses. The discrepancy may be understood
with the additional kinetic energy \cite{Park:2015nha}. From the comparison for results in the current model \cite{Wu:2016vtq} and in a
dynamical study \cite{Liu:2019zuc}, the calculated masses of heavy-full tetraquark states are truly higher than the highest masses
estimated with meson-meson thresholds but lower than the theoretical
upper limits. In the following discussions, we use
this feature as a criterion for reasonable tetraquark masses. The reasonability of the results may be tested in future
studies.

\subsection{Effective interactions and supplemental results for the $cs\bar{c}\bar{s}$ system}

In Ref. \cite{Wu:2016vtq}, we have discussed the effects on the
tetraquark masses due to change of coupling parameters and argued
the stability of $QQ\bar{Q}\bar{Q}$ states by using the effective
color-spin interactions in the case that the mixing of different
color-spin structures is considered. In Ref. \cite{Li:2018vhp}, we
further introduced a dimensionless measure to reflect the effective
color-spin interaction between the $i$th quark component and the
$j$th quark component,
\begin{eqnarray}
K_{ij}=\frac{\Delta M}{\Delta C_{ij}}\to \frac{\partial M}{\partial
C_{ij}}.
\end{eqnarray}
With such measures, one may rewrite the multiquark masses as
\begin{eqnarray}\label{eftKij}
M=M_0+\sum_{i<j}K_{ij}C_{ij}.
\end{eqnarray}
When $K_{ij}$ is a negative (positive) number, the effective
interaction between the $i$th and $j$th quark components is
attractive (repulsive). If $K_{12}$ and $K_{34}$ are negative but $K_{13}$,
$K_{14}$, $K_{23}$, and $K_{24}$ are positive, the tetraquark state
$Q_1q_2\bar{Q}_3\bar{q}_4$ is probably more stable than other cases.
If only $K_{12}$ or $K_{34}$ is negative, the state is probably less
stable than the mentioned case but more stable than other cases. In the following parts, we qualitatively discuss the stability of tetraquarks with such effective interactions.

\begin{table}[!h]
 \caption{$K_{ij}$'s for $cs\bar{c}\bar{s}$ states. The order of states for each case of $J^{PC}$ is the same as that of masses from high to low.}\label{cscscncn-Kij}
\begin{tabular}{ccccc}\hline
&\multicolumn{4}{c}{$cs\bar{c}\bar{s}$ system} \\ \hline \hline
$J^{PC}$ &  $K_{cs}$ & $K_{c\bar{c}}$ & $K_{c\bar{s}}$ & $K_{s\bar{s}}$  \\
$2^{++}$&$\left[\begin{array}{c}-2.1\\4.7\end{array}\right]$&$\left[\begin{array}{c}4.7\\0.0\end{array}\right]$&$\left[\begin{array}{c}3.4\\5.9\end{array}\right]$&$\left[\begin{array}{c}4.7\\0.0\end{array}\right]$\\
$1^{++}$&$\left[\begin{array}{c}-0.4\\-2.3\end{array}\right]$&$\left[\begin{array}{c}5.3\\-0.7\end{array}\right]$&$\left[\begin{array}{c}0.3\\-9.7\end{array}\right]$&$\left[\begin{array}{c}5.3\\-0.7\end{array}\right]$\\
$1^{+-}$&$\left[\begin{array}{c}-0.1\\3.8\\-1.7\\-1.9\end{array}\right]$&$\left[\begin{array}{c}-1.0\\-5.2\\-8.1\\5.0\end{array}\right]$&$\left[\begin{array}{c}10.4\\-1.5\\-6.3\\-2.6\end{array}\right]$&$\left[\begin{array}{c}1.1\\1.2\\2.9\\-14.6\end{array}\right]$\\
$0^{++}$&$\left[\begin{array}{c}7.0\\-12.1\\6.0\\-6.2\end{array}\right]$&$\left[\begin{array}{c}3.9\\2.6\\-4.6\\-11.2\end{array}\right]$&$\left[\begin{array}{c}7.2\\4.9\\-16.5\\-14.2\end{array}\right]$&$\left[\begin{array}{c}3.9\\2.6\\-4.6\\-11.2\end{array}\right]$\\
\hline
\end{tabular}
\begin{tabular}{ccccc}\hline
&\multicolumn{4}{c}{$cn\bar{c}\bar{n}$ system} \\ \hline \hline
$J^{PC}$ & $K_{cn}$ & $K_{c\bar{c}}$ & $K_{c\bar{n}}$ & $K_{n\bar{n}}$  \\
$2^{++}$&$\left[\begin{array}{c}-0.5\\3.2\end{array}\right]$&$\left[\begin{array}{c}5.3\\-0.6\end{array}\right]$&$\left[\begin{array}{c}0.6\\8.7\end{array}\right]$&$\left[\begin{array}{c}5.3\\-0.6\end{array}\right]$\\
$1^{++}$&$\left[\begin{array}{c}-0.3\\-2.4\end{array}\right]$&$\left[\begin{array}{c}5.3\\-0.7\end{array}\right]$&$\left[\begin{array}{c}0.3\\-9.6\end{array}\right]$&$\left[\begin{array}{c}5.3\\-0.7\end{array}\right]$\\
$1^{+-}$&$\left[\begin{array}{c}2.4\\-3.8\\2.1\\-0.7\end{array}\right]$&$\left[\begin{array}{c}-7.1\\-8.1\\0.6\\5.3\end{array}\right]$&$\left[\begin{array}{c}7.3\\1.8\\-8.4\\-0.7\end{array}\right]$&$\left[\begin{array}{c}3.2\\3.1\\0.2\\-15.9\end{array}\right]$\\
$0^{++}$&$\left[\begin{array}{c}5.7\\-10.3\\1.6\\-2.4\end{array}\right]$&$\left[\begin{array}{c}4.6\\2.3\\-0.7\\-15.5\end{array}\right]$&$\left[\begin{array}{c}5.2\\4.9\\-25.8\\-3.0\end{array}\right]$&$\left[\begin{array}{c}4.6\\2.3\\-0.7\\-15.5\end{array}\right]$\\
\hline
\end{tabular}
\end{table}

In ours previous work \cite{Wu:2016gas}, we considered the spectrum of $cs\bar{c}\bar{s}$ states. Here, we do not repeat the results given there, but present the supplemental results about effective interactions. The obtained $K_{ij}$'s of Eq. \eqref{eftKij} are listed in Table \ref{cscscncn-Kij}. The order of states for each case of $J^{PC}$ is the same as the order of masses from high to low. From the results, the highest $2^{++}$, the highest $1^{++}$, and the second highest $0^{++}$ states are
probably more stable than other states. Although the $X(4274)$ as another $1^{++}$ $cs\bar{c}\bar{s}$ state is higher than the $X(4140)$, its width can be narrower than that of $X(4140)$. This feature is not contradicted with the recent LHCb measurement \cite{Aaij:2016iza}.

\subsection{The $cn\bar{c}\bar{n}$,  $bn\bar{b}\bar{n}$, and $bs\bar{b}\bar{s}$ systems}\label{sec3.1}

These three systems have similar structures to the $cs\bar{c}\bar{s}$ case but involve different
values of parameters. They are related to most of the
charmonium-like or bottomonium-like $XYZ$ states observed in various
processes. The quantum numbers of the tetraquark states may be
$J^{PC}=2^{++}$, $1^{++}$, $1^{+-}$, or $0^{++}$. In the literature,
there are lots of studies relevant with such tetraquarks, see e.g. Refs.
\cite{Chao:1979mm,Chao:1979tg,Ebert:2005nc,Cui:2006mp,Zhang:2007xa,Vijande:2007fc,Chen:2017dpy,Yang:2017rmm,Anwar:2018sol}. Here, one gets the eigenvalues of the chromomagnetic interactions by substituting $\tau=2C_{Qq}$, $\theta=0$,
$\alpha=C_{Q\bar{Q}}+C_{q\bar{q}}+2C_{Q\bar{q}}$,
$\beta=\mu=C_{Q\bar{Q}}-C_{q\bar{q}}$, and
$\nu=C_{Q\bar{Q}}+C_{q\bar{q}}-2C_{Q\bar{q}}$ into Eqs.
\eqref{matrixJ2} and \eqref{matrixJ0} and by diagonalizing matrices
in Eqs. \eqref{matrixJ2}, \eqref{matrixJ0}, \eqref{matrixJ1pp}, and
\eqref{matrixJ1pm}. For the $Qn\bar{Q}\bar{n}$ systems, the
isospin=1 and isospin=0 states are both allowed. In the current
model, the obtained isovector and isoscalar $Qn\bar{Q}\bar{n}$ states
are degenerate.

\begin{table}[!h]
\caption{Calculated CMI eigenvalues and estimated tetraquark masses
for the $cn\bar{c}\bar{n}$ systems in units
of MeV. The masses in the fourth column (Upper limits) are obtained
with Eq. \eqref{mass} and those in the last three columns with
various reference states.}\label{cscscncn}
\begin{tabular}{c|cccccccccccccc}\hline\hline
\multicolumn{7}{c}{$cn\bar{c}\bar{n}$ system} \\ \hline
$J^{PC}$ &$\langle H_{CM}\rangle$ &Eigenvalues& Upper limits & $\eta_c \pi$ & $D\bar{D}$ & X(4140)   \\ 
$2^{++}$&$\left(\begin{array}{cc}86.0&-61.4\\-61.4&151.0\end{array}\right)$&$\left(\begin{array}{c}188.0\\49.0\end{array}\right)$&$\left(\begin{array}{c}4361\\4222\end{array}\right)$&$\left(\begin{array}{c}3873\\3734\end{array}\right)$&$\left(\begin{array}{c}4132\\3993\end{array}\right)$&$\left(\begin{array}{c}4237\\4098\end{array}\right)$\\
$1^{++}$&$\left(\begin{array}{cc}7.6&-137.2\\-137.2&83.0\end{array}\right)$&$\left(\begin{array}{c}187.6\\-97.0\end{array}\right)$&$\left(\begin{array}{c}4360\\4076\end{array}\right)$&$\left(\begin{array}{c}3873\\3588\end{array}\right)$&$\left(\begin{array}{c}4132\\3847\end{array}\right)$&$\left(\begin{array}{c}4236\\3952\end{array}\right)$\\
$1^{+-}$&$\left(\begin{array}{cccc}-43.3&61.4&-65.3&138.6\\61.4&-172.3&138.6&-163.3\\-65.3&138.6&-50.3&137.2\\138.6&-163.3&137.2&-61.7\end{array}\right)$&$\left(\begin{array}{c}116.4\\47.0\\-38.8\\-452.3\end{array}\right)$&$\left(\begin{array}{c}4289\\4220\\4134\\3720\end{array}\right)$&$\left(\begin{array}{c}3802\\3732\\3646\\3233\end{array}\right)$&$\left(\begin{array}{c}4060\\3991\\3905\\3492\end{array}\right)$&$\left(\begin{array}{c}4165\\4096\\4010\\3597\end{array}\right)$\\
$0^{++}$&$\left(\begin{array}{cccc}-108.0&122.8&-50.1&237.6\\122.8&-334.0&237.6&-125.3\\-50.1&237.6&-64.0&0.0\\237.6&-125.3&0.0&32.0\end{array}\right)$&$\left(\begin{array}{c}221.0\\71.8\\-192.7\\-574.1\end{array}\right)$&$\left(\begin{array}{c}4394\\4244\\3980\\3598\end{array}\right)$&$\left(\begin{array}{c}3906\\3757\\3492\\3111\end{array}\right)$&$\left(\begin{array}{c}4165\\4016\\3751\\3370\end{array}\right)$&$\left(\begin{array}{c}4270\\4121\\3856\\3475\end{array}\right)$\\
\hline
\end{tabular}
\end{table}

First, we consider the $cn\bar{c}\bar{n}$ system. When Eq.
\eqref{mass} is adopted, one gets the theoretically highest
tetraquark masses listed in the fourth column of Table \ref{cscscncn} . When the reference system is chosen as
$\eta_c\pi$, the lowest masses in our strategy are obtained and listed in the fifth column of Table \ref{cscscncn}. When
the reference mass is chosen as the $D\bar{D}$ threshold, one gets
more reasonable masses shown in the sixth column of Table \ref{cscscncn}, but they are probably still lower than the
realistic values. If there were one meson that we may identify as a
compact $cn\bar{c}\bar{n}$ tetraquark, the relatively reliable
masses of its partner states may be estimated with the CMI
eigenvalues. However, as argued in Sec. \ref{sec1}, it is hard for
us to identify such a tetraquark state. To perform a better mass
estimation, an alternative method we may try is to relate the
$cn\bar{c}\bar{n}$ masses to that of the $X(4140)$. To do that, we
rewrite the mass formula Eq. \eqref{mass} as
\begin{eqnarray}
M_{(cn\bar{c}\bar{n})}&=&2m_c+2m_n+\langle
H_{CM}\rangle_{(cn\bar{c}\bar{n})}=(2m_c+2m_s)-2(m_s-m_n)+\langle
H_{CM}\rangle_{(cn\bar{c}\bar{n})}.
\end{eqnarray}
For the $(2m_c+2m_s)$ term, we replace it by $M_{X(4140)}-\langle
H_{CM}\rangle_{X(4140)}$. Then the additional attraction that $m_c$
(=1724.6 MeV) and $m_s$ (=540.3 MeV) should incorporate is partly
compensated. For the $(m_s-m_n)$ (=178.6 MeV) term, we also need a
modification. Now the problem of mass estimation becomes the problem
to determine mass gap between different quarks. The extracted
$(m_s-m_n)$ from the conventional hadrons varies from 90.8 MeV to
187.1 MeV (see Table \ref{quarkmassdifference}). If one replaces the
larger value 178.6 MeV by the smaller value $m_s-m_n=90.8$ MeV,
i.e., the mass formula in estimating the $cn\bar{c}\bar{n}$ masses
is
\begin{eqnarray}
M_{(cn\bar{c}\bar{n})}&=&M_{X(4140)}-2m_{B_s}+2m_B+2\langle
H_{CM}\rangle_{B_s}-2\langle H_{CM}\rangle_{B}-\langle
H_{CM}\rangle_{X(4140)}+\langle H_{CM}\rangle_{(cn\bar{c}\bar{n})},
\end{eqnarray}
higher masses than those with the $D\bar{D}$ threshold are obtained
(see Table \ref{cscscncn}). If one uses the larger value
$m_s-m_n=187.1$ MeV, the obtained tetraquark masses are
2$\times$(187.1-90.8)=192.6 MeV lower than those in the last column.
Then the masses are not reasonable according to the above criterion.
Considering the quark environment, probably the quark mass
difference between $m_s$ in the $cs\bar{c}\bar{s}$ system and $m_n$
in the $cn\bar{c}\bar{n}$ system is close to that between $D_s$ and
$D$. If this is the case, the $cn\bar{c}\bar{n}$ masses are just
25.4 MeV lower than those in the last column. In the following, we
assume that the masses in the last column of Table \ref{cscscncn}
are reasonable values. Of course, further studies are required to
test this method of mass estimation.

\begin{table}[!h]
\caption{Quark mass differences (units: MeV) determined with various
hadrons. The values from the extracted effective quark masses are
$m_s-m_n$= 178.6 MeV and $m_b-m_c$=3328.2
MeV.}\label{quarkmassdifference} \centering
\begin{tabular}{ccc|ccc}\hline\hline
Hadron&Hadron&$(m_s-m_n)$ & Hadron&Hadron&$(m_b-m_c)$ \\
\hline\hline
$D_s$&$D$&103.5&   $B$&$D$&3340.9\\
$B_s$&$B$&90.8 &   $B_s$&$D_s$&3328.2\\
$\Sigma$&$N$&187.1 &   $\eta_b$&$\eta_c$&3188.4\\
$\Lambda$&$N$&177.4&   $\Lambda_b$&$\Lambda_c$&3333.1\\
$\Omega_c$&$\Sigma_c$&158.8&  $\Sigma_b$&$\Sigma_c$&3328.5\\
$\Omega_b$&$\Sigma_b$&147.9&  $\Xi_b$&$\Xi_c$&3326.2\\
$\Xi_c$&$\Lambda_c$&133.4  &  $\Omega_b$&$\Omega_c$&3315.7\\
$\Xi_c$&$\Sigma_c$&119.5\\
$\Xi_b$&$\Lambda_b$&126.9\\
$\Xi_b$&$\Sigma_b$&117.6\\
\hline
\end{tabular}
\end{table}

\begin{figure}[htpb]
\includegraphics[width=0.6\textwidth]{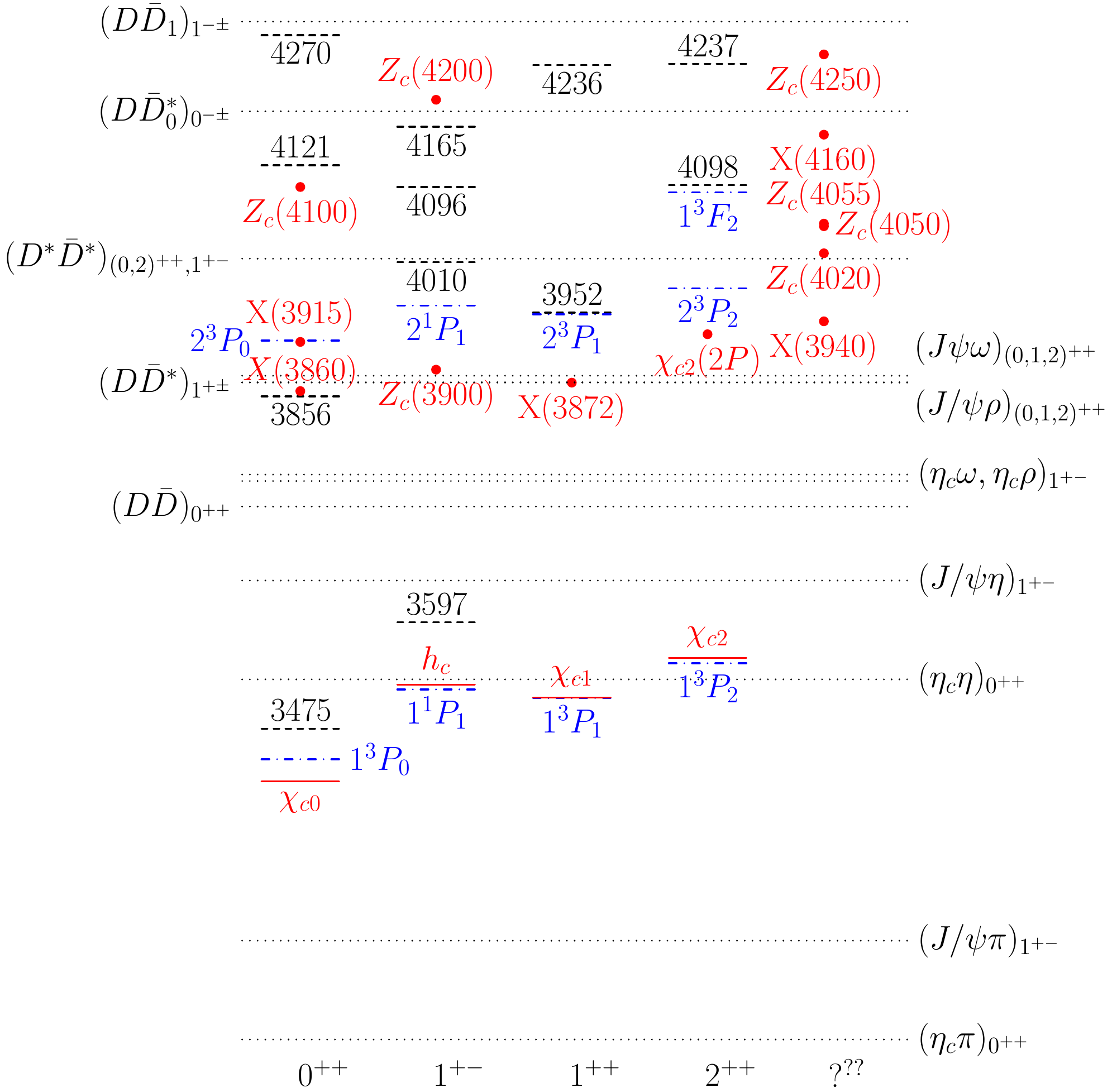}
\caption{Relative positions for $cn\bar{c}\bar{n}$ tetraquarks
(black dashed lines), predicted charmonia (blue dash-dotted lines),
observed charmonia (red solid lines), states with exotic properties
(red solid dots), and various meson-meson thresholds (black dotted
lines). The masses are given in units of MeV. The subscripts of
threshold symbols are $J^{PC}$ in the $S$-wave
case.}\label{fig-cncn}
\end{figure}

One should note that the $cn\bar{c}\bar{n}$ masses in Table \ref{cscscncn} are both for isovector and for isoscalar tetraquark states. The important mixing effects for all the quantum numbers are not small. In Fig. \ref{fig-cncn}, we show the relative positions for the $cn\bar{c}\bar{n}$ tetraquark states, predicted QM charmonia \cite{Godfrey:1985xj}, relevant observed states, and various meson-meson thresholds. For the meson-mesons channels, we label their $S$-wave $J^{PC}$ in the subscripts of their symbols. It is convenient to judge whether a state can decay into a meson-meson channel from the $J$, $P$, and $C$ conservations or not. In Table \ref{cscscncn-Kij}, we also show the obtained $K_{ij}$'s of Eq. \eqref{eftKij} for the $cn\bar{c}\bar{n}$ states from which one may guess relatively stable tetraquarks.

With the help of the relative positions in Fig. \ref{fig-cncn}, one may discuss possible assignments for the exotic charmonium-like mesons shown in the figure. For convenience, we summarize the mesons we will discuss, their quantum numbers, masses, widths, and finding channels in Table \ref{expXYZ}. In the particle data book \cite{Tanabashi:2018oca}, the $Z_c(3900)$ and $Z_c(3885)$ are assumed as the same state and $Z_c(4020)$ and $Z_c(4025)$ are treated as the same state. Here, we also adopt such assignments.

\begin{table}[!h]
\caption{Properties of mesons related with ground $cn\bar{c}\bar{n}$
tetraquark states \cite{Tanabashi:2018oca}.}\label{expXYZ}
\centering
\begin{tabular}{ccccccc}
\hline\hline
States&$I^G(J^{PC})$&Mass (MeV)&Width (MeV)&Finding channels\\
$X(3860)$&$0^+(0^{++})$&$3862^{+26+40}_{-32-13}$&$201^{+154+88}_{-67-82}$& Belle: $D\bar{D}$ \cite{Chilikin:2017evr}\\
$X(3872)$&$0^+(1^{++})$&$3871.69\pm0.17$&$<1.2$&Belle: $J\psi\pi\pi$ \cite{Choi:2003ue}\\
$X(3915)$&$0^+(0/2^{++})$&$3918.4\pm1.9$&$20\pm5$&Belle: $J/\psi\omega$ \cite{Abe:2004zs}\\
$X(3940)$&$?^?(?^{??})$&$3942^{+7}_{-6}\pm6$&$37^{+26}_{-15}\pm8$&Belle: $D\bar{D}^*$ \cite{Abe:2007sya}\\
$X(4160)$&$?^?(?^{??})$&$4156^{+25}_{-20}\pm15$&$139^{+111}_{-61}\pm21$&Belle: $D^*\bar{D}^*$ \cite{Abe:2007sya}\\
$Z_c(3900)$&$1^+(1^{+-})$&$3886.6\pm2.4$&$28.2\pm2.6$&BESIII: $J/\psi\pi$ \cite{Ablikim:2013mio},$D\bar{D}^*$ \cite{Ablikim:2013xfr}\\
$Z_c(4020)$&$1^+(?^{?-})$&$4024.1\pm1.9$&$13\pm5$&BESIII: $h_c\pi$ \cite{Ablikim:2013wzq}, $D^*\bar{D}^*$ \cite{Ablikim:2013emm}\\
$Z_c(4050)$&$1^-(?^{?+})$&$4051\pm14^{+20}_{-41}$&$82^{+21+47}_{-17-22}$&Belle: $\chi_{c1}\pi$ \cite{Mizuk:2008me}\\
$Z_c(4055)$&$1^+(?^{?-})$&$4054\pm3\pm1$&$45\pm11\pm6$& Belle: $\psi(2S)\pi$ \cite{Wang:2014hta}\\
$Z_c(4100)$&$1^?(0^{+?}/1^{-?})$&$4096\pm20^{+18}_{-22}$&$152\pm58^{+60}_{-35}$& LHCb: $\eta_c\pi$ \cite{Aaij:2018bla}\\
$Z_c(4200)$&$1^+(1^{+-})$&$4196^{+31+17}_{-29-13}$&$370\pm70^{+70}_{-132}$&Belle: $J/\psi\pi$ \cite{Chilikin:2014bkk}\\
$Z_c(4250)$&$1^-(?^{?+})$&$4248^{+44+180}_{-29-35}$&$177^{+54+316}_{-39-61}$&Belle: $\chi_{c1}\pi$ \cite{Mizuk:2008me}\\
$Z_c(4430)$&$1^+(1^{+-})$&$4478^{+15}_{-18}$&$181\pm31$&Belle: $\psi(2S)\pi$ \cite{Choi:2007wga}\\
\hline
\end{tabular}
\end{table}
We start the discussions with the newly observed charged
$Z_c(4100)^-$. Its quark content should be $cd\bar{c}\bar{u}$. From
Fig. \ref{fig-cncn}, this state ($J^P=0^+$ or $1^-$) is $\sim80$ MeV above the threshold
of $D^*D^*$ and $\sim190$ MeV below the threshold of $D\bar{D}_1$. It is unlikely an $S$- or $P$-wave meson-meson state, but definite conclusion needs detailed investigations. Our results indicate that the second highest $J^{PC}=0^{++}$ $cnc\bar{n}$ tetraquark (the $C$-parity of the
neutral partner is +) has a mass close to that of $Z_c(4100)$. One
may interpret the $Z_c(4100)$ as a scalar tetraquark state. By
comparing relative positions for the state and thresholds in Fig.
\ref{fig-cncn}, the $Z_c(4100)$ can decay into $D^*\bar{D}^*$,
$D\bar{D}$, $J/\psi\rho$, and $\eta_c\pi$ through $S$-wave
interactions if it is really a state. Its width should be very broad
and the state may be even unobservable. If we check Table
\ref{cscscncn-Kij}, from $K_{cn}=-10.3$, the color-spin interaction
between the charm quark and the light quark is effectively
attractive. Although the color-spin interactions between other quark
components are effectively repulsive, the coefficient for the $cn$
interaction is larger than those for others. This means that the
state has a relatively stable tetraquark structure. The observed
$\Gamma=152$ MeV for this high mass state $Z_c(4100)$ is also
qualitatively consistent with the argument that it is a scalar
tetraquark. In fact, in a study with the QCD sum rule (QSR)
\cite{Chen:2017dpy}, the calculation also indicates that a $0^{++}$
tetraquark around 4.1 GeV is possible. If the $J^P$ of $Z_c(4100)$
are $1^-$, the $J^{PC}$ of its neutral partner will be $1^{-+}$. In
the QSR calculation, such a tetraquark has a mass around 4.6 GeV
\cite{Chen:2010ze}, which means that the $Z_c(4100)$ is more like a
scalar state. In addition, Ref. \cite{Wang:2018ntv} fails to
reproduce the mass of the $Z_c(4100)$ with an interpolating current
for vector tetraquarks. All these results favor the $0^{++}$
tetraquark assignment for the $Z_c(4100)$. If this is a correct
interpretation, the state may also be observed in the $J/\psi\rho$,
$D\bar{D}$, and $D^*\bar{D}^*$ channels. Because of the degeneracy
of isovector and isoscalar tetraquarks in the present model, an
isoscalar state around 4.1 GeV is also possible. Experimentally, it
can be searched for in the $J/\psi\omega$ and $\eta_c\eta$ channels.

If the above assignment is correct, probably the $X(3860)$ is another $0^{++}$ tetraquark. This state was observed in the $D\bar{D}$ channel at Belle \cite{Chilikin:2017evr} and the $J^{PC}=0^{++}$ assignment is more favored than $2^{++}$. From our estimation, the second lowest $0^{++}$ tetraquark is close to it, which is a signal that the $X(3860)$ is probably a $cn\bar{c}\bar{n}$ tetraquark state. The QSR calculation also gives an isoscalar scalar tetraquark around 3.81 GeV \cite{Chen:2017dpy} which is consistent with the $X(3860)$. Both the mass and width of a scalar tetraquark consistent with the $X(3860)$ are obtained in another QSR investigation \cite{Wang:2017lbl}. If we check the amplitudes for the effective color-spin interactions in Table \ref{cscscncn-Kij}, one finds that the $c\bar{n}$ interaction for the tetraquark is stronger and this state should not be stable like the second highest tetraquark. The resulting width may be comparable to that of the $Z_c(4100)$ although it is below the $Z_c(4100)$. The observed width $\sim201$ MeV for the $X(3860)$ is qualitatively consistent with this feature. In Ref. \cite{Chilikin:2017evr}, the $X(3860)$ was interpreted as the $\chi_{c0}(2P)$ because the $\chi_{c0}(2P)$ should decay dominantly into $D\bar{D}$ while the $X(3915)$ which was once identified as $\chi_{c0}(2P)$ does not satisfy this requirement. However, the analysis in Ref. \cite{Zhou:2017dwj} indicates that the $\chi_{c0}(2P)$ has a narrow width, which is inconsistent with Belle's result. The $X(3860)$ is unlikely to be a charmonium from its broad width. In order to understand its nature, $\eta_c\eta$ and $\eta_c\pi$ channels are proposed to search for such a state and its isospin partner state, respectively. 

In the $0^{++}$ case, four additional tetraquark states, two around
3470 MeV ($I=1$ and $I=0$) and two around 4270 MeV ($I=1$ and
$I=0$), are also possible. The low mass $cn\bar{c}\bar{n}$
tetraquark states are around the predicted $\chi_{c0}$. The only
$S$-wave rearrangement decay mode for the isovector state is
$\eta_c\pi$ and no rearrangement decay modes exist for the isoscalar
one. Their widths may not be broad if they do exist. We wait for
experimental measurements to test this judgement. The high mass
$cn\bar{c}\bar{n}$ states around 4.2 GeV should be broader than the
$Z_c(4100)$. Experimentally, there is a charged $Z_c(4250)$ in this
mass region, which is observed in the $\pi\chi_{c1}$ channel and has
a width around 177 MeV. It is unlikely a $D_1\bar{D}$ or
$D_0\bar{D}^*$ molecule \cite{Ding:2008gr}. At present, assigning
the $Z_c(4250)$ as a ground tetraquark with highest mass can not be
excluded. If this assignment is correct, an additional isoscalar
tetraquark around 4.2 GeV should also be measurable. However, since
the quantum numbers of $Z_c(4250)$ can also be $1^{++}$, $2^{++}$,
or others and the highest tetraquarks in the $1^{++}$ and $2^{++}$
cases are also in this mass region, there are still other
possibilities for its assignment.

The $\chi_{c0}$ charmonium has been established, but its radially
excited $\chi_{c0}(2P)$ not. This state should be around 3920 MeV.
Experimentally, three narrow states in this mass region, $X(3915)$
in the $J/\psi\omega$ channel
\cite{Abe:2004zs,Uehara:2009tx,delAmoSanchez:2010jr,Lees:2012xs},
$\chi_{c2}(2P)$ in the $D\bar{D}$ channel
\cite{Uehara:2005qd,Aubert:2010ab}, and $X(3940)$ in the
$D\bar{D}^*$ channel \cite{Abe:2007sya}, were observed. The angular
momentum of the state observed in the $D\bar{D}$ channel has been
determined to be 2 and this state is identified as the predicted
$2^3P_2$ charmonium. For the other two states, the assignment
problem is still unsettled. From Refs.
\cite{Zhou:2017dwj,Zhou:2015uva}, the $X(3915)$ and the state in
$D\bar{D}$ are probably the same $2^{++}$ state while the real
$\chi_{c0}(2P)$ is probably around 3860 MeV with a narrow width. For
the $X(3940)$ state, there is no appropriate position if it is a
$P$-even charmonium ($X(3872)$ should be the $\chi_{c1}(2P)$ state).
From a study of the decay width in Ref. \cite{Wang:2016mqb}, this
exotic state seems to be a good candidate of $\eta_c(3S)$. If the
above assignments for the $Z_c(4100)$ and $X(3860)$ are correct, the
widths of the $cn\bar{c}\bar{n}$ tetraquark states should not be
small. Since $\Gamma_{X(3940)}\sim 37$ MeV, the consistency of decay
widths does not support its tetraquark interpretation. Based on our
results, it seems that no tetraquark assignments are favored for
states around 3940 MeV. These three or two states around 3920 MeV
are probably conventional charmonia or molecules.

Now let us move on to the $J^{PC}=2^{++}$ states. In the charmonium
sector, the lowest $\chi_{c2}$ has been established and the
$\chi_{c2}(2P)$ is also identified. No evidence for other charmonia
has been reported. In the tetraquark sector, we have two states
around 4.1 GeV ($I=1$ and $I=0$) and two states around 4.2 GeV
($I=1$ and $I=0$). Their dominant decay channels should be
$D^*\bar{D}^*$. For the isovector (isoscalar) states, the decay mode
$J/\psi\rho$ ($J/\psi\omega$) is also allowed. Whether such
tetraquark states exist or not needs to be answered by future
measurements. As mentioned above, the $Z_c(4250)$ can also be a
candidate of the high mass $2^{++}$ tetraquark. From Table
\ref{cscscncn-Kij}, the $cn$ effective color-spin interaction in
this state is weakly attractive, which probably narrows its width.

In the $J^{PC}=1^{++}$ case, the most intriguing state is $X(3872)$
which is probably the predicted $\chi_{c1}(2P)$ charmonium but
affected strongly by channels coupling to it. The lowest $1^{++}$
tetraquark we obtain is also around the predicted $\chi_{c1}(2P)$
charmonium which is tens of MeV above the physical $X(3872)$. This
indicates that the $X(3872)$ should not be a pure tetraquark, which
is consistent with the results in Refs.
\cite{Zhang:2007xa,Vijande:2007fc}. If the coupling between the
predicted charmonium and the isoscalar $cn\bar{c}\bar{n}$ tetraquark
state is considered, it is possible to obtain the physical mass of
the $X(3872)$. On the other hand, the isovector $cn\bar{c}\bar{n}$,
in principle, does not couple to conventional charmonium states. Its
dominant decay modes are $D\bar{D}^*$ and $J/\psi\rho$. If
experiments observed an isovector state around 3950 MeV (with
probably broad width), it will be a good tetraquark candidate.
Around 4.2 GeV, we have two higher tetraquarks ($I=1$ and $I=0$).
The dominant decay modes for the isovector state are still
$D\bar{D}^*$ and $J/\psi\rho$. Those for the isoscalar are
$D\bar{D}^*$ and $J/\psi\omega$. The large phase spaces for decay
indicate that both of them should be broad if they exist. Note that
probably the $Z_c(4250)$ can also be the high mass $1^{++}$
tetraquark whose effective $cn$ interaction is weakly attractive.

The remaining quantum numbers that the ground tetraquarks involve
are $J^{PC}=1^{+-}$. The exotic $1^{+-}$ $Z_c(3900)$ states have
been observed around the $D\bar{D}^*$ threshold
\cite{Ablikim:2013mio,Liu:2013dau,Xiao:2013iha}. From our estimated
masses, one cannot interpret the $Z_c(3900)$ states as ground
compact tetraquarks. They should be isovector $D\bar{D}^*$ bound or
resonant states, or non-resonant effects, as explored in the
literature
\cite{Chen:2016qju,Albaladejo:2016jsg,Gong:2016jzb,Ikeda:2016zwx,Ikeda:2017mee,He:2017lhy,Ortega:2018cnm}.
Another state consistent with $J^{PC}=1^{+-}$ is $Z_c(4200)$
observed in the $J/\psi\pi$ channel. Its width is about 370 MeV,
which implies that this state is probably a tetraquark. From our
results of estimation, one may assign it as the highest ground
$cn\bar{c}\bar{n}$ tetraquark. In QSR analyses
\cite{Chen:2015fsa,Wang:2015nwa}, the tetraquark assignment for the
$Z_c(4200)$ is also favored. In this mass region, there is an exotic
$X(4160)$ which was observed in the $D^*\bar{D}^*$ channel by Belle
\cite{Abe:2007sya}. Although its width ($\sim139$ MeV) is broad,
assigning it as a $1^{+-}$ tetraquark seems to be problematic
because (1) its mass is larger than the $Z_c(4100)$, (2) it has more
decay channels than the $Z_c(4100)$, and (3) the $cn$ diquark is not
effectively attractive, but the width is not larger than the
$Z_c(4100)$. Possible assignments may be $\eta_{c2}(2D)$
\cite{Yang:2009fj}, $\eta_c(4S)$ \cite{Wang:2016mqb}, or
$D_s^*\bar{D}_s^*$ molecule with $I^G(J^{PC})=0^+(2^{++})$
\cite{Molina:2009ct}. The $Z_c(4430)$ observed in the $\pi\psi(2S)$
channel \cite{Chilikin:2013tch} also has the quantum numbers
$J^{PC}=1^{+-}$. It is much higher than the $Z_c(4200)$ and should
be an excited state. We do not discuss its nature here. Three
charged tetraquark states, one around 4.1 GeV, the other around 4.0
GeV, and the third around 3.6 GeV, are also possible. The lowest one
has only one rearrangement decay mode $J/\psi\pi$ and is probably
not a broad tetraquark. The other two should be broad. For the
isoscalar $1^{+-}$ tetraquark states, there is still no candidate we
can assign. The $J/\psi\eta$ may be an ideal channel to identify
them because the decay of the conventional $1^{+-}$ charmonia into
$J/\psi$ involves spin-flip and is suppressed. The lowest $I=0$
tetraquark (around 3.6 GeV) should be very narrow if it really
exists.

In the above discussions, we do not mention the three charged states
in the mass range 4.0$\sim$4.1 GeV, $Z_c(4055)$, $Z_c(4050)$, and
$Z_c(4020)$, although we need experimental candidates to assign as
tetraquarks. These three states do not have broad enough widths that
consistent assignments for ground tetraquark states require. Their
nature should be accounted for by other interpretations. For
example, the $Z_c(4020)$ can be explained as coupled channel cusp
effect \cite{Swanson:2014tra} or $1^{+-}$ molecule-like state
\cite{Chen:2013omd,Cui:2013vfa,Patel:2014vua,Chen:2015jwa} and the
$Z_c(4050)$ as $3^{++}$ molecule-like state \cite{Patel:2014vua}.
They can also be radially or orbitally excited four-quark states,
which has not been widely studied in the literature
\cite{Chen:2016qju}.

From the symmetry consideration, its tetraquark partners should also
exist if the $Z_c(4100)$ is really a $0^{++}$ $cn\bar{c}\bar{n}$
tetraquark state. Quite a few broad isovector and isoscalr exotic
states can be searched for with the help of Fig. \ref{fig-cncn}.
Four low mass (and probably narrow) $cn\bar{c}\bar{n}$ tetraquarks
are also possible. If such additional states could be observed, we
will be sure that more compact tetraquark states in other systems
exist. The corresponding state of $Z_c(4100)$ in the
$cs\bar{c}\bar{s}$ case (mass$\sim$4.2 GeV) should also be
relatively stable and can be observed since the $cs$ color-spin
interaction is also effectively attractive. Further experimental
measurements are definitely needed.

\begin{table}[htbp]
\caption{Calculated CMI eigenvalues and estimated tetraquark masses
for the $bn\bar{b}\bar{n}$ and $bs\bar{b}\bar{s}$ systems in units
of MeV. The masses in the forth column (Upper limits) are obtained
with Eq. \eqref{mass} and those in the last three columns with
various reference states.}\label{bnbnbsbs}
\begin{tabular}{c|cccccccccccc}\hline
\multicolumn{7}{c}{$bn\bar{b}\bar{n}$ system} \\ \hline \hline
$J^{PC}$ &$\langle H_{CM}\rangle$&  Eigenvalues & Upper limits
&$\eta_b \pi$ & $B\bar{B}$ & X(4140)   \\ \hline
$2^{++}$&$\left(\begin{array}{cc}56.1&-80.6\\-80.6&119.5\end{array}\right)$&$\left(\begin{array}{c}174.5\\1.2\end{array}\right)$&$\left(\begin{array}{c}11003\\10830\end{array}\right)$&$\left(\begin{array}{c}10236\\10063\end{array}\right)$&$\left(\begin{array}{c}10800\\10627\end{array}\right)$&$\left(\begin{array}{c}10905\\10732\end{array}\right)$\\
$1^{++}$&$\left(\begin{array}{cc}31.1&-104.4\\-104.4&98.5\end{array}\right)$&$\left(\begin{array}{c}174.4\\-44.9\end{array}\right)$&$\left(\begin{array}{c}11003\\10784\end{array}\right)$&$\left(\begin{array}{c}10236\\10017\end{array}\right)$&$\left(\begin{array}{c}10800\\10581\end{array}\right)$&$\left(\begin{array}{c}10905\\10686\end{array}\right)$\\
$1^{+-}$&$\left(\begin{array}{cccc}-42.3&80.6&-71.7&152.2\\80.6&-126.5&152.2&-179.3\\-71.7&152.2&-44.9&104.4\\152.2&-179.3&104.4&-91.5\end{array}\right)$&$\left(\begin{array}{c}115.5\\58.3\\-17.2\\-461.7\end{array}\right)$&$\left(\begin{array}{c}10944\\10887\\10812\\10367\end{array}\right)$&$\left(\begin{array}{c}10177\\10120\\10045\\9600\end{array}\right)$&$\left(\begin{array}{c}10741\\10684\\10609\\10164\end{array}\right)$&$\left(\begin{array}{c}10846\\10789\\10713\\10269\end{array}\right)$\\
$0^{++}$&$\left(\begin{array}{cccc}-91.5&161.2&-65.8&180.8\\161.2&-249.5&180.8&-164.5\\-65.8&180.8&-20.8&0.0\\180.8&-164.5&0.0&10.4\end{array}\right)$&$\left(\begin{array}{c}178.8\\67.1\\-72.9\\-524.3\end{array}\right)$&$\left(\begin{array}{c}11008\\10896\\10756\\10305\end{array}\right)$&$\left(\begin{array}{c}10241\\10129\\9989\\9538\end{array}\right)$&$\left(\begin{array}{c}10805\\10693\\10553\\10102\end{array}\right)$&$\left(\begin{array}{c}10909\\10798\\10658\\10206\end{array}\right)$\\
\hline\hline \multicolumn{7}{c}{$bs\bar{b}\bar{s}$ system} \\ \hline
$J^{PC}$ &$\langle H_{CM}\rangle$ &  Eigenvalues& Upper limits & $\eta_b\phi$ & $B_s\bar{B_s}$ & X(4140)   \\ 
$2^{++}$&$\left(\begin{array}{cc}30.4&-25.0\\-25.0&56.9\end{array}\right)$&$\left(\begin{array}{c}71.9\\15.4\end{array}\right)$&$\left(\begin{array}{c}11258\\11202\end{array}\right)$&$\left(\begin{array}{c}10481\\10424\end{array}\right)$&$\left(\begin{array}{c}10879\\10823\end{array}\right)$&$\left(\begin{array}{c}10984\\10928\end{array}\right)$\\
$1^{++}$&$\left(\begin{array}{cc}5.4&-51.0\\-51.0&32.6\end{array}\right)$&$\left(\begin{array}{c}71.8\\-33.8\end{array}\right)$&$\left(\begin{array}{c}11258\\11152\end{array}\right)$&$\left(\begin{array}{c}10481\\10375\end{array}\right)$&$\left(\begin{array}{c}10879\\10774\end{array}\right)$&$\left(\begin{array}{c}10984\\10879\end{array}\right)$\\
$1^{+-}$&$\left(\begin{array}{cccc}-17.6&25.0&-20.3&43.1\\25.0&-63.3&43.1&-50.8\\-20.3&43.1&-18.2&51.0\\43.1&-50.8&51.0&-26.2\end{array}\right)$&$\left(\begin{array}{c}35.3\\6.8\\-13.4\\-154.1\end{array}\right)$&$\left(\begin{array}{c}11222\\11193\\11173\\11032\end{array}\right)$&$\left(\begin{array}{c}10444\\10416\\10395\\10255\end{array}\right)$&$\left(\begin{array}{c}10843\\10814\\10794\\10653\end{array}\right)$&$\left(\begin{array}{c}10948\\10919\\10899\\10758\end{array}\right)$\\
$0^{++}$&$\left(\begin{array}{cccc}-41.7&49.9&-20.4&88.3\\49.9&-123.4&88.3&-51.0\\-20.4&88.3&-19.2&0.0\\88.3&-51.0&0.0&9.6\end{array}\right)$&$\left(\begin{array}{c}81.4\\29.4\\-67.3\\-218.2\end{array}\right)$&$\left(\begin{array}{c}11268\\11216\\11119\\10968\end{array}\right)$&$\left(\begin{array}{c}10490\\10438\\10341\\10191\end{array}\right)$&$\left(\begin{array}{c}10889\\10837\\10740\\10589\end{array}\right)$&$\left(\begin{array}{c}10994\\10942\\10845\\10694\end{array}\right)$\\
\hline
\end{tabular}
\end{table}

\begin{figure}[htpb]
\begin{tabular}{ccc}
\includegraphics[width=220pt]{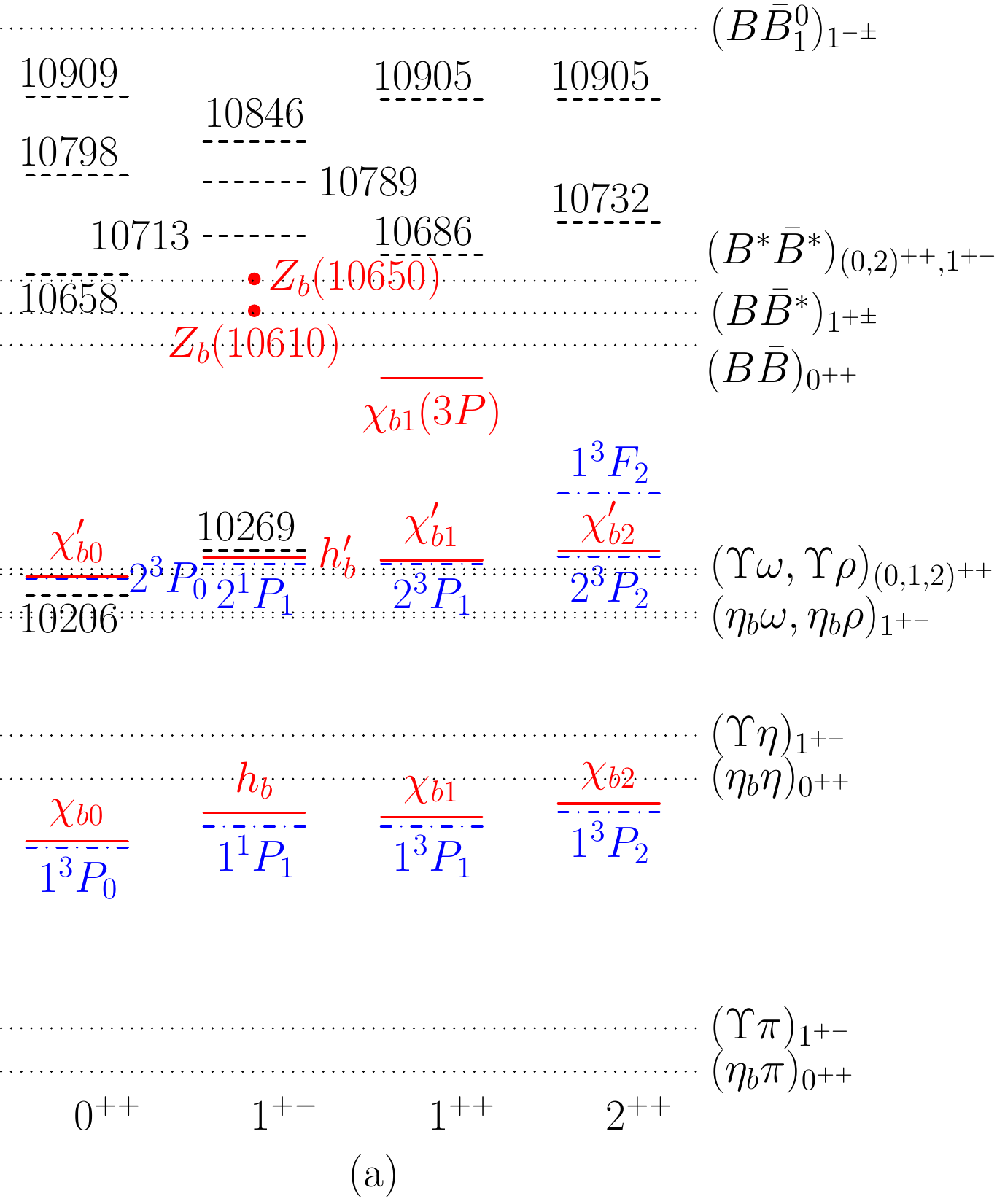}&$\quad$&
\includegraphics[width=220pt]{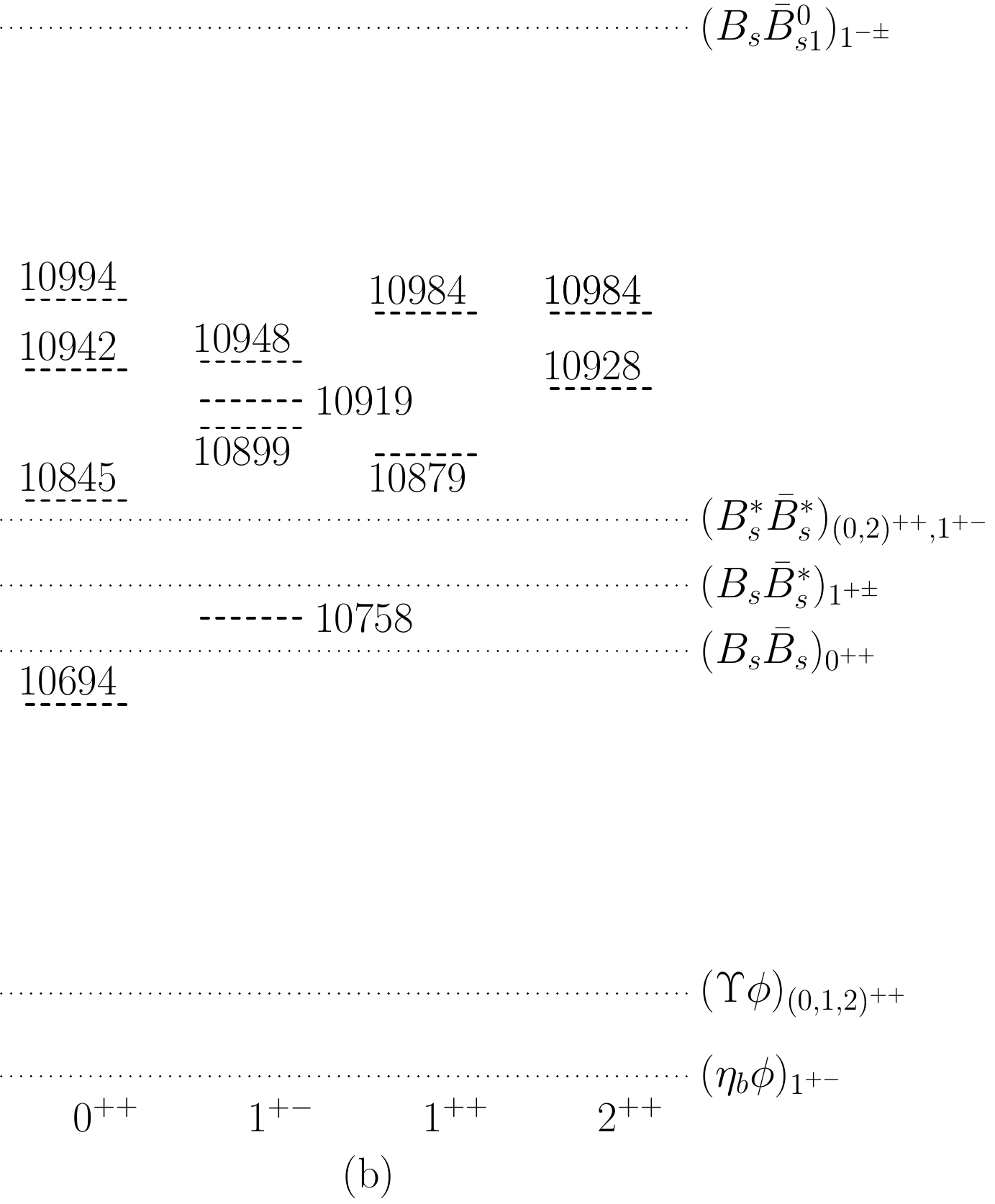}
\end{tabular}
\caption{Relative positions for (a) $bn\bar{b}\bar{n}$ and (b)
$bs\bar{b}\bar{s}$ tetraquarks (black dashed lines), predicted
bottomonia (blue dash-dotted lines), observed bottomonia (red solid
lines), states with exotic properties (red solid dots), and various
meson-meson thresholds (black dotted lines). The masses are given in
units of MeV. The subscripts of threshold symbols are $J^{PC}$ in
the $S$-wave case.}\label{fig-bnbnbsbs}
\end{figure}

Secondly, we consider the $bn\bar{b}\bar{n}$ system. The obtained CMI
eigenvalues, theoretical upper limits for the tetraquark masses, and
estimated values with the $\eta_b\pi$ and $B\bar{B}$ thresholds are
listed in Table \ref{bnbnbsbs}. Similar to the estimation procedure
for the $cn\bar{c}\bar{n}$ states, when one relates the masses to
that of the $X(4140)$, we rewrite the mass formula to be
\begin{eqnarray}
M_{(bn\bar{b}\bar{n})}&=&2m_b+2m_n+\langle
H_{CM}\rangle_{(bn\bar{b}\bar{n})}=(2m_c+2m_s)+2(m_b-m_c)-2(m_s-m_n)+\langle
H_{CM}\rangle_{(bn\bar{b}\bar{n})}.
\end{eqnarray}
Then the $(2m_c+2m_s)$ term is replaced by $M_{X(4140)}-\langle
H_{CM}\rangle_{X(4140)}$ and the $(m_s-m_n)$ term is replaced by
90.8 MeV. For the mass difference $(m_b-m_c)$, there are some
uncertainties with typical values of tens of MeV (see Table
\ref{quarkmassdifference}). By using 3340.9 MeV extracted from $B$
and $D$ mesons, one may get higher masses than those determined with
the $B\bar{B}$ threshold. If the value from the $\eta_b$ and
$\eta_c$ is adopted, tetraquark masses lower than those determined
with the $B\bar{B}$ threshold are obtained. They should not be
reasonable values according to our criterion. In the following
discussions, we assume the masses listed in the seventh column are
closer to the realistic values. With such masses, we plot in Fig.
\ref{fig-bnbnbsbs} relative positions for the $bn\bar{b}\bar{n}$
tetraquarks, predicted QM bottomonia, relevant observed states, and
various meson-meson thresholds. Unlike the $cn\bar{c}\bar{n}$ case,
only two narrow exotic bottomonium-like states, $Z_b(10650)$ and
$Z_b(10610)$ \cite{Belle:2011aa}, were observed in the present case.
To understand effective color-spin interactions in the tetraquark
states, in Table \ref{bnbnbsbs-Kij}, we give the involved $K_{ij}$'s
for the $bn\bar{b}\bar{n}$ states.

\begin{table}[htbp]
 \caption{$K_{ij}$'s for $bn\bar{b}\bar{n}$ and $bs\bar{b}\bar{s}$ states. The order of states is the same as that in Table \ref{bnbnbsbs}.}\label{bnbnbsbs-Kij}
\begin{tabular}{ccccc}\hline
&\multicolumn{4}{c}{$bn\bar{b}\bar{n}$ system} \\ \hline \hline
$J^{PC}$&$K_{bn}$ & $K_{b\bar{b}}$ & $K_{b\bar{n}}$ & $K_{n\bar{n}}$  \\
$2^{++}$&$\left[\begin{array}{c}-0.1\\2.8\end{array}\right]$&$\left[\begin{array}{c}5.3\\-0.7\end{array}\right]$&$\left[\begin{array}{c}0.1\\9.2\end{array}\right]$&$\left[\begin{array}{c}5.3\\-0.7\end{array}\right]$\\
$1^{++}$&$\left[\begin{array}{c}-0.1\\-2.6\end{array}\right]$&$\left[\begin{array}{c}5.3\\-0.7\end{array}\right]$&$\left[\begin{array}{c}0.1\\-9.4\end{array}\right]$&$\left[\begin{array}{c}5.3\\-0.7\end{array}\right]$\\
$1^{+-}$&$\left[\begin{array}{c}1.6\\-2.7\\1.3\\-0.2\end{array}\right]$&$\left[\begin{array}{c}-15.2\\-1.3\\1.9\\5.3\end{array}\right]$&$\left[\begin{array}{c}1.8\\2.2\\-3.7\\-0.2\end{array}\right]$&$\left[\begin{array}{c}5.2\\2.0\\-0.6\\-16.0\end{array}\right]$\\
$0^{++}$&$\left[\begin{array}{c}2.4\\-5.0\\-2.1\\-0.6\end{array}\right]$&$\left[\begin{array}{c}5.2\\2.0\\-0.6\\-16.0\end{array}\right]$&$\left[\begin{array}{c}2.5\\3.9\\-24.4\\-0.7\end{array}\right]$&$\left[\begin{array}{c}5.2\\2.0\\-0.6\\-16.0\end{array}\right]$\\
\hline
\end{tabular}
\begin{tabular}{ccccc}\hline
&\multicolumn{4}{c}{$bs\bar{b}\bar{s}$ system} \\ \hline \hline
$J^{PC}$& $K_{bs}$ & $K_{b\bar{b}}$ & $K_{b\bar{s}}$ & $K_{s\bar{s}}$  \\
$2^{++}$&$\left[\begin{array}{c}-0.5\\3.2\end{array}\right]$&$\left[\begin{array}{c}5.3\\-0.6\end{array}\right]$&$\left[\begin{array}{c}0.6\\8.7\end{array}\right]$&$\left[\begin{array}{c}5.3\\-0.6\end{array}\right]$\\
$1^{++}$&$\left[\begin{array}{c}-0.3\\-2.4\end{array}\right]$&$\left[\begin{array}{c}5.3\\-0.7\end{array}\right]$&$\left[\begin{array}{c}0.3\\-9.6\end{array}\right]$&$\left[\begin{array}{c}5.3\\-0.7\end{array}\right]$\\
$1^{+-}$&$\left[\begin{array}{c}0.8\\-1.5\\1.3\\-0.7\end{array}\right]$&$\left[\begin{array}{c}-2.9\\-11.1\\-0.6\\5.3\end{array}\right]$&$\left[\begin{array}{c}8.9\\0.7\\-8.9\\-0.7\end{array}\right]$&$\left[\begin{array}{c}2.1\\3.7\\0.7\\-15.9\end{array}\right]$\\
$0^{++}$&$\left[\begin{array}{c}5.2\\-9.7\\1.1\\-2.0\end{array}\right]$&$\left[\begin{array}{c}4.8\\2.2\\-0.6\\-15.7\end{array}\right]$&$\left[\begin{array}{c}4.8\\5.2\\-26.3\\-2.4\end{array}\right]$&$\left[\begin{array}{c}4.8\\2.2\\-0.6\\-15.7\end{array}\right]$\\
\hline
\end{tabular}
\end{table}

From comparison for Figs. \ref{fig-cncn} and \ref{fig-bnbnbsbs}(a),
the mass distribution for the $bn\bar{b}\bar{n}$ system is similar
to that for $cn\bar{c}\bar{n}$. Figure \ref{fig-bnbnbsbs} tells us
that most $bn\bar{b}\bar{n}$ tetraquarks have open-bottom decay
channels and should be broad. The lowest $0^{++}$ tetraquark with
$I=1$ ($I=0$) mainly decays into $\eta_b\pi$ ($\eta_b\eta$) through
$S$-wave interactions. The lowest $1^{+-}$ tetraquark with $I=1$
($I=0$) mainly decays into $\eta_b\rho$ and $\Upsilon\pi$
($\eta_b\omega$ and $\Upsilon\eta$). Maybe they are not broad
states. From Table \ref{bnbnbsbs-Kij}, the $bn$ interaction in the
second highest $I^G(J^{PC})=1^-(0^{++})$ tetraquark, the
corresponding state of $Z_c(4100)$ in the hidden-bottom case, is
also effectively attractive. It should also be a measurable broad
state. So does its degenerate $I=0$ partner state (mass around 10.8
GeV). Similarly, the highest $2^{++}$ and $1^{++}$
$bn\bar{b}\bar{n}$ tetraquarks are probably measurable since their
effective $bn$ interactions are weakly attractive while the
effective interactions for other quark components in them are
repulsive.

From the discussions in the $cn\bar{c}\bar{n}$ case, it seems that
tetraquark states generally have broad widths. By changing the
(anti)charm quark to the (anti)bottom quark, this basic feature of
tetraquark states probably does not change. The observed two $Z_b$
states have narrow widths, $\Gamma_{Z_b(10610)}=18.4$ MeV and
$\Gamma_{Z_b(10650)}=11.5$ MeV \cite{Belle:2011aa}, and they should
not be compact tetraquarks. Since they are near-threshold states,
the natural explanation is that they are hadronic molecules
\cite{Bondar:2011ev,Sun:2011uh,Wang:2018jlv}. The $Z_c(3900)$ near
the $D\bar{D}^*$ threshold has a width around 28 MeV and is
basically thought as a $1^{+-}$ $D\bar{D}^*$ molecule. As its
hidden-bottom partner, the $Z_b(10610)$ is a $1^{+-}$ $B\bar{B}^*$
molecule. If the $Z_b(10650)$ is a $1^{+-}$ $B^*\bar{B}^*$ molecule,
the $Z_c(4020)$ with $\Gamma\sim13$ MeV looks like its hidden-charm
partner and the $J^{PC}$ of $Z_c(4020)$ should be $1^{+-}$, too. The
assignment for the $J^{PC}$ from this simple comparison may be
tested with future measurements.

Searching for more exotic states in the hidden-bottom realm is an
intriguing task since the bottom and charm quarks have different
properties. The observation of them will be crucial for us to
understand the quark interactions in conventional hadrons and in
multiquark states, no matter the observed width is broad or narrow.
We hope the results in the present work may provide useful
information for further studies.

Finally, we consider the $bs\bar{b}\bar{s}$ system. We present
relevant masses in Table \ref{bnbnbsbs} and various $K_{ij}$'s in
Table \ref{bnbnbsbs-Kij}. When relating the masses to that of the
$X(4140)$, we modify terms in
\begin{eqnarray}
M_{(bs\bar{b}\bar{s})}&=&2m_b+2m_s+\langle
H_{CM}\rangle_{(bs\bar{b}\bar{s})}=(2m_c+2m_s)+2(m_b-m_c)+\langle
H_{CM}\rangle_{(bs\bar{b}\bar{s})}.
\end{eqnarray}
By replacing $(2m_c+2m_s)$ with $M_{X(4140)}-\langle
H_{CN}\rangle_{X(4140)}$ and $(m_b-m_c)$ with 3340.9 MeV, we get
$bs\bar{b}\bar{s}$ masses larger than those determined with the
$B_s\bar{B}_s$ threshold. They are shown in the last column of Table
\ref{bnbnbsbs}. We treat them as the realistic masses and plot
relative tetraquark positions in Fig. \ref{fig-bnbnbsbs} (b).
Relevant rearrangement decay channels and their thresholds are also
shown.

From the figure, these $bs\bar{b}\bar{s}$ tetraquarks can be
searched for either in $\eta_b\phi$ or $\Upsilon\phi$ channel. All
of them seem to have open-bottom decay channels, which is a feature
different from the $cs\bar{c}\bar{s}$ case \cite{Wu:2016gas}.
However, it is unclear whether they are broad or narrow states
because the $X(4140)$ as a tetraquark has a narrow width around 22
MeV \cite{Tanabashi:2018oca}. We hope future investigations may
answer this puzzle. From Table \ref{bnbnbsbs-Kij}, the highest
$2^{++}$, $1^{++}$, and the second highest $0^{++}$ states seem to
be more stable than other states since the effective $cs$ color-spin
interactions are attractive.

\subsection{The $cn\bar{c}\bar{s}$, $bn\bar{b}\bar{s}$, $cn\bar{b}\bar{n}$, and $cs\bar{b}\bar{s}$ systems}\label{sec3.2}

The existence of isovector charmonium-like and bottomonium-like
tetraquark states also implies that of more exotic tetraquarks. One
may find some predictions about the $cn\bar{c}\bar{s}$,
$bn\bar{b}\bar{s}$, $cn\bar{b}\bar{n}$, and $cs\bar{b}\bar{s}$
states in Refs.
\cite{Ebert:2005nc,Cui:2006mp,Chen:2013aba,Agaev:2016dsg,Agaev:2017uky}.
The $cn\bar{c}\bar{s}$ and $bn\bar{b}\bar{s}$ states look like
excited kaon mesons from the quantum numbers but the masses are much higher. If such a high-mass kaon were observed, one may identify its tetraquark nature since the orbital or
radial excitation energy larger than 3 GeV for light quarks in a conventional kaon is unlikely. The creation of a heavy quark-antiquark pair can naturally explain its high mass. The $cn\bar{b}\bar{n}$ and $cs\bar{b}\bar{s}$ states look
like excited $B_c$ mesons, but probably they are not easy to be
isolated from the conventional $B_c$ mesons. All such tetraquark
states do not have $C$-parities. To get numerical results in the
present model, the matrices in Eqs. \eqref{matrixJ2},
\eqref{matrixJ0}, and \eqref{matrixJ1} need to be diagonalized after
appropriate parameters are used.

\begin{figure}[htbp]
\begin{tabular}{ccc}
\includegraphics[width=220pt]{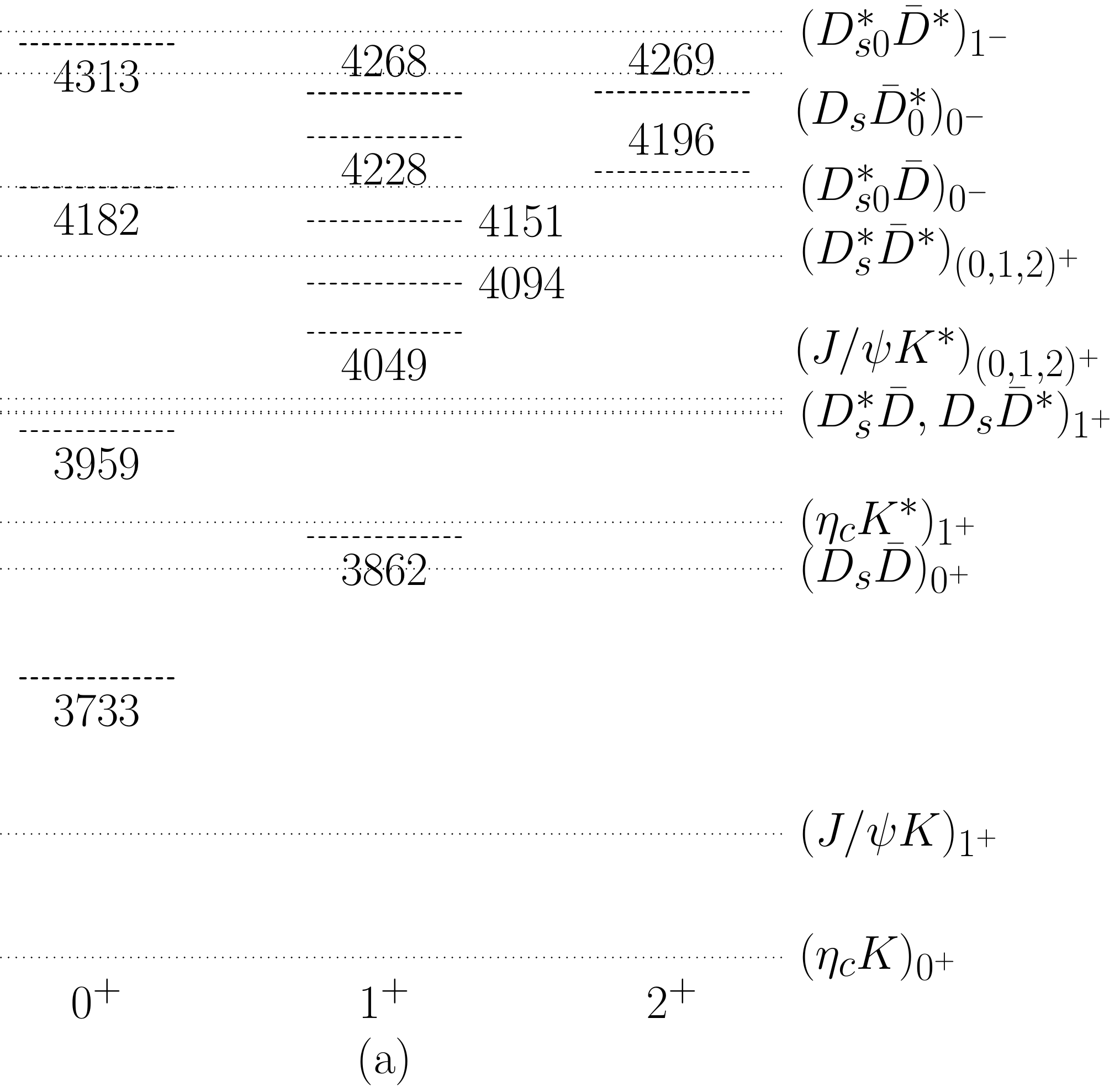}&\quad&
\includegraphics[width=220pt]{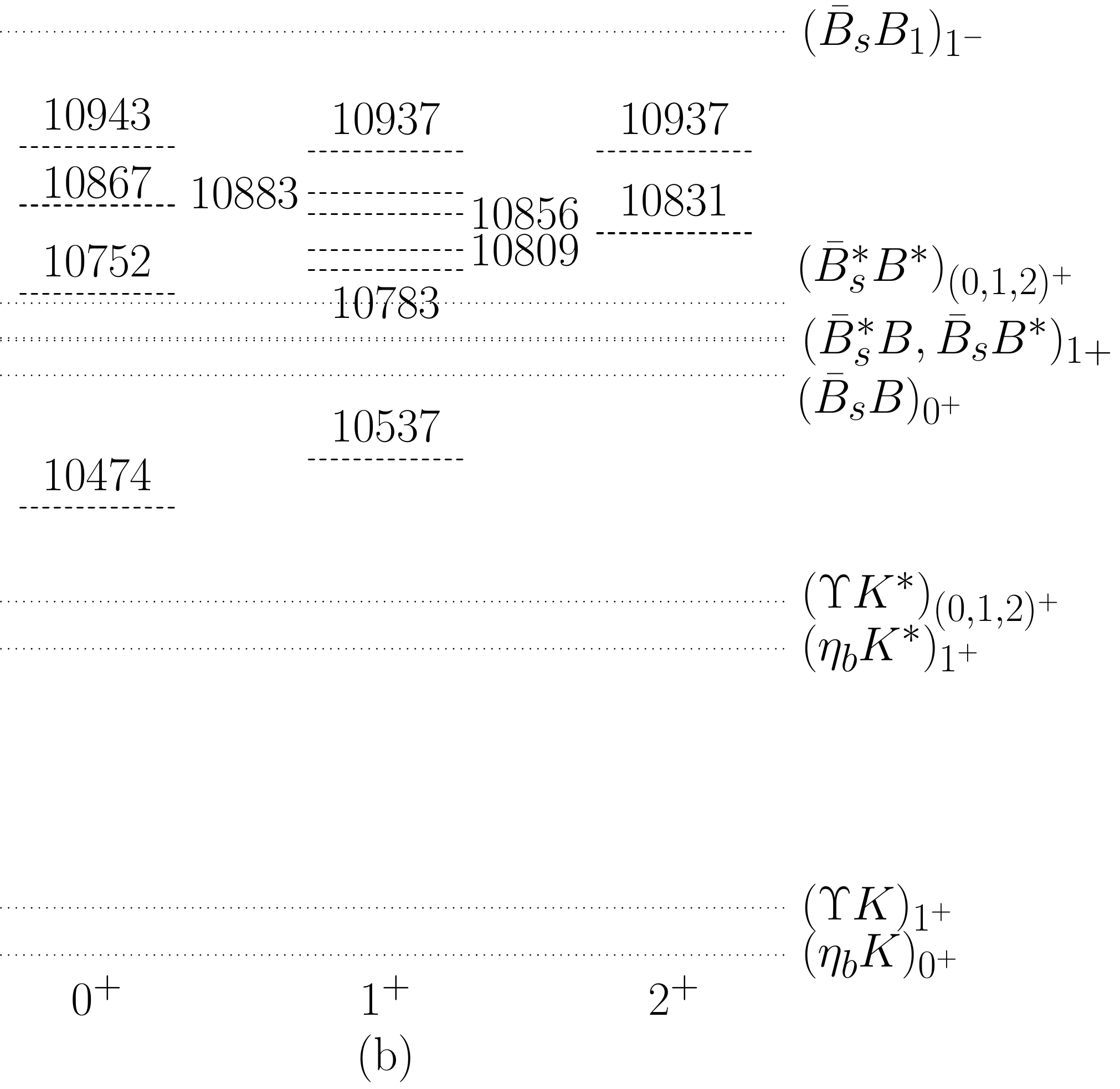}
\end{tabular}
\caption{Relative positions for kaon-like heavy tetraquarks (black
dashed lines): (a) $cn\bar{c}\bar{s}$ and (b) $bn\bar{b}\bar{s}$ and
various meson-meson thresholds (black dotted lines). The masses are
given in units of MeV. The subscripts of threshold symbols are
$J^{PC}$ in the $S$-wave case.}\label{fig-cncsbnbs}
\end{figure}

We consider temporarily the kaon-like heavy tetraquark states. With
Eqs. \eqref{matrixJ2}, \eqref{matrixJ0}, and \eqref{matrixJ1}, the
numerical results for the chromomagnetic interactions can be easily
gotten. We list them in Table \ref{cncsbnbs}. The theoretical upper
limits for the tetraquark masses, the masses estimated with the
$\eta_cK$ ($\eta_bK$) threshold, and those with the $\bar{D}D_s$
($B\bar{B}_s$) threshold are given in the forth, fifth, and sixth
columns, respectively. The masses with $\bar{D}D_s$ ($B\bar{B}_s$)
are higher than those with $\eta_cK$ ($\eta_bK$). To estimate the
masses with the help of $X(4140)$, we adopt modified mass formulas
of
\begin{eqnarray}
M_{(cn\bar{c}\bar{s})}&=&2m_c+m_n+m_s+\langle
H_{CM}\rangle_{(cn\bar{c}\bar{s})}=(2m_c+2m_s)-(m_s-m_n)+\langle
H_{CM}\rangle_{(cn\bar{c}\bar{s})},
\end{eqnarray}
and
\begin{eqnarray}
M_{(bn\bar{b}\bar{s})}&=&2m_b+m_n+m_s+\langle
H_{CM}\rangle_{(bn\bar{b}\bar{s})}=(2m_c+2m_s)+2(m_b-m_c)-(m_s-m_n)+\langle
H_{CM}\rangle_{(bn\bar{b}\bar{s})}.
\end{eqnarray}
By making the replacements $(2m_c+2m_s)\to M_{X(4140)}-\langle
H_{CM}\rangle_{X(4140)}$, $(m_b-m_c)\to 3340.9$ MeV, and
$(m_s-m_n)\to 90.8$ MeV, we obtain much higher masses in the last
column of Table \ref{cncsbnbs}. We treat them as more reasonable
values in the following discussions. The relative positions for the
kaon-like heavy tetraquark states and relevant meson-meson channels
and thresholds are illustrated in Fig. \ref{fig-cncsbnbs}.
Contributions of effective quark interactions for each pair of quark
components are easy to recover with Eq. \eqref{eftKij} and the
coefficients $K_{ij}$'s in Table \ref{cncsbnbs-Kij}.

\begin{table}[htbp]
\caption{Calculated CMI eigenvalues and estimated tetraquark masses
for the $cn\bar{c}\bar{s}$ and $bn\bar{b}\bar{s}$ systems in units
of MeV. The masses in the forth column (Upper limits) are obtained
with Eq. \eqref{mass} and those in the last three columns with
various reference states.}\label{cncsbnbs}
\begin{tabular}{c|cccccccccccc}\hline
&\multicolumn{6}{c}{$cn\bar{c}\bar{s}$ system} \\ \hline \hline
$J^{P}$ &$\langle H_{CM}\rangle$&  Eigenvalues& Upper limits &
$\eta_c K$ & $\bar{D}D_s$ & X(4140)   \\ \hline
$2^{+}$&$\left(\begin{array}{cc}72.5&-30.0\\-30.0&113.3\end{array}\right)$&$\left(\begin{array}{c}129.2\\56.7\end{array}\right)$&$\left(\begin{array}{c}4480\\4408\end{array}\right)$&$\left(\begin{array}{c}3991\\3918\end{array}\right)$&$\left(\begin{array}{c}4177\\4104\end{array}\right)$&$\left(\begin{array}{c}4269\\4196\end{array}\right)$\\
$1^{+}$&$\left(\begin{array}{cccccc}-27.2&-25.3&25.3&30.0&53.6&-53.6\\-25.3&-25.3&14.1&53.6&0.0&-105.8\\25.3&14.1&-20.0&-53.6&-105.8&0.0\\30.0&53.6&-53.6&-136.0&-63.2&63.2\\53.6&0.0&-105.8&-63.2&12.7&35.3\\-53.6&-105.8&0.0&63.2&35.3&10.0\end{array}\right)$&$\left(\begin{array}{c}128.5\\87.9\\10.9\\-45.5\\-90.3\\-277.3\end{array}\right)$&$\left(\begin{array}{c}4480\\4439\\4362\\4306\\4261\\4074\end{array}\right)$&$\left(\begin{array}{c}3990\\3950\\3872\\3816\\3771\\3584\end{array}\right)$&$\left(\begin{array}{c}4176\\4135\\4058\\4002\\3957\\3770\end{array}\right)$&$\left(\begin{array}{c}4268\\4228\\4151\\4094\\4049\\3862\end{array}\right)$\\
$0^{+}$&$\left(\begin{array}{cccc}-77.1&60.0&-24.5&183.2\\60.0&-260.7&183.2&-61.2\\-24.5&183.2&-68.0&0.0\\183.2&-61.2&0.0&34.0\end{array}\right)$&$\left(\begin{array}{c}172.9\\42.7\\-180.5\\-406.9\end{array}\right)$&$\left(\begin{array}{c}4524\\4394\\4171\\3944\end{array}\right)$&$\left(\begin{array}{c}4035\\3904\\3681\\3455\end{array}\right)$&$\left(\begin{array}{c}4220\\4090\\3867\\3641\end{array}\right)$&$\left(\begin{array}{c}4313\\4182\\3959\\3733\end{array}\right)$\\
\hline\hline
&\multicolumn{6}{c}{$bn\bar{b}\bar{s}$ system} \\ \hline
$J^{P}$ &$\langle H_{CM}\rangle$&  Eigenvalues& Upper limits & $\eta_b K$& $B\bar{B}_s$ & X(4140)   \\ 
$2^{+}$&$\left(\begin{array}{cc}41.3&-48.6\\-48.6&83.3\end{array}\right)$&$\left(\begin{array}{c}115.3\\9.3\end{array}\right)$&$\left(\begin{array}{c}11123\\11017\end{array}\right)$&$\left(\begin{array}{c}10354\\10248\end{array}\right)$&$\left(\begin{array}{c}10832\\10726\end{array}\right)$&$\left(\begin{array}{c}10937\\10831\end{array}\right)$\\
$1^{+}$&$\left(\begin{array}{cccccc}-28.0&-30.2&29.4&48.6&62.4&-64.0\\-30.2&-6.1&22.9&62.4&0.0&-73.5\\29.4&22.9&-7.2&-64.0&-73.5&0.0\\48.6&62.4&-64.0&-90.0&-75.4&73.5\\62.4&0.0&-73.5&-75.4&3.1&57.3\\-64.0&-73.5&0.0&73.5&57.3&3.6\end{array}\right)$&$\left(\begin{array}{c}115.3\\61.2\\34.1\\-12.4\\-38.5\\-284.3\end{array}\right)$&$\left(\begin{array}{c}11123\\11069\\11042\\10995\\10969\\10723\end{array}\right)$&$\left(\begin{array}{c}10354\\10300\\10272\\10226\\10200\\9954\end{array}\right)$&$\left(\begin{array}{c}10832\\10778\\10751\\10704\\10678\\10432\end{array}\right)$&$\left(\begin{array}{c}10937\\10883\\10856\\10809\\10783\\10537\end{array}\right)$\\
$0^{+}$&$\left(\begin{array}{cccc}-62.7&97.3&-39.7&127.4\\97.3&-176.7&127.4&-99.3\\-39.7&127.4&-20.0&0.0\\127.4&-99.3&0.0&10.0\end{array}\right)$&$\left(\begin{array}{c}121.7\\45.5\\-69.2\\-347.4\end{array}\right)$&$\left(\begin{array}{c}11129\\11053\\10938\\10660\end{array}\right)$&$\left(\begin{array}{c}10360\\10284\\10169\\9891\end{array}\right)$&$\left(\begin{array}{c}10838\\10762\\10647\\10369\end{array}\right)$&$\left(\begin{array}{c}10943\\10867\\10752\\10474\end{array}\right)$\\
\hline
\end{tabular}
\end{table}

\begin{table}[!h]
\caption{$K_{ij}$'s for $cn\bar{c}\bar{s}$ and $bn\bar{b}\bar{s}$
states. The order of states is the same as that in Table
\ref{cncsbnbs}.}\label{cncsbnbs-Kij}\tiny
\begin{tabular}{c|cccccc}\hline
\multicolumn{7}{c}{$cn\bar{c}\bar{s}$ system} \\ \hline \hline
$J^{P}$ & $K_{cn}$ & $K_{c\bar{c}}$ & $K_{c\bar{s}}$ & $K_{c\bar{n}}$ & $K_{s\bar{n}}$ & $K_{cs}$ \\
$2^{+}$&$\left[\begin{array}{c}-0.5\\1.8\end{array}\right]$&$\left[\begin{array}{c}5.2\\-0.6\end{array}\right]$&$\left[\begin{array}{c}0.6\\4.1\end{array}\right]$&$\left[\begin{array}{c}0.6\\4.1\end{array}\right]$&$\left[\begin{array}{c}5.2\\-0.6\end{array}\right]$&$\left[\begin{array}{c}-0.5\\1.8\end{array}\right]$\\
$1^{+}$&$\left[\begin{array}{c}-0.5\\0.7\\-0.2\\0.3\\-1.1\\-0.5\end{array}\right]$&$\left[\begin{array}{c}5.3\\-3.0\\-9.2\\-2.4\\-0.7\\5.3\end{array}\right]$&$\left[\begin{array}{c}0.1\\4.7\\0.3\\-4.2\\-5.1\\-0.5\end{array}\right]$&$\left[\begin{array}{c}0.2\\4.6\\0.5\\-4.8\\-4.5\\-0.7\end{array}\right]$&$\left[\begin{array}{c}5.3\\2.0\\3.0\\1.3\\-0.7\\-15.6\end{array}\right]$&$\left[\begin{array}{c}0.1\\0.1\\-0.1\\0.4\\-1.3\\-0.6\end{array}\right]$\\
$0^{+}$&$\left[\begin{array}{c}3.3\\-5.8\\1.8\\-2.0\end{array}\right]$&$\left[\begin{array}{c}4.3\\2.5\\-1.6\\-14.5\end{array}\right]$&$\left[\begin{array}{c}3.1\\2.4\\-11.8\\-3.1\end{array}\right]$&$\left[\begin{array}{c}3.1\\2.4\\-11.8\\-3.1\end{array}\right]$&$\left[\begin{array}{c}4.3\\2.5\\-1.6\\-14.5\end{array}\right]$&$\left[\begin{array}{c}3.3\\-5.8\\1.8\\-2.0\end{array}\right]$\\
\hline
\end{tabular}
\begin{tabular}{c|cccccc}\hline
\multicolumn{7}{c}{$bn\bar{b}\bar{s}$ system} \\ \hline \hline
$J^{P}$ &  $K_{bn}$ & $K_{b\bar{b}}$ & $K_{b\bar{s}}$ & $K_{b\bar{n}}$ & $K_{s\bar{n}}$ & $K_{bs}$ \\
$2^{+}$&$\left[\begin{array}{c}-0.1\\1.5\end{array}\right]$&$\left[\begin{array}{c}5.3\\-0.7\end{array}\right]$&$\left[\begin{array}{c}0.1\\4.5\end{array}\right]$&$\left[\begin{array}{c}0.1\\4.5\end{array}\right]$&$\left[\begin{array}{c}5.3\\-0.7\end{array}\right]$&$\left[\begin{array}{c}-0.1\\1.5\end{array}\right]$\\
$1^{+}$&$\left[\begin{array}{c}-0.0\\1.4\\-2.3\\1.0\\-1.3\\-0.2\end{array}\right]$&$\left[\begin{array}{c}5.3\\-11.6\\-4.6\\1.6\\-0.7\\5.3\end{array}\right]$&$\left[\begin{array}{c}0.1\\2.3\\1.0\\-1.8\\-6.0\\-0.2\end{array}\right]$&$\left[\begin{array}{c}0.0\\2.3\\0.8\\-4.2\\-3.5\\-0.2\end{array}\right]$&$\left[\begin{array}{c}5.3\\4.3\\2.6\\-0.3\\-0.7\\-16.0\end{array}\right]$&$\left[\begin{array}{c}-0.1\\1.5\\-2.1\\0.8\\-1.2\\-0.2\end{array}\right]$\\
$0^{+}$&$\left[\begin{array}{c}1.8\\-3.5\\-0.4\\-0.5\end{array}\right]$&$\left[\begin{array}{c}5.1\\2.0\\-0.5\\-15.9\end{array}\right]$&$\left[\begin{array}{c}1.7\\2.4\\-12.9\\-0.6\end{array}\right]$&$\left[\begin{array}{c}1.7\\2.4\\-12.9\\-0.6\end{array}\right]$&$\left[\begin{array}{c}5.1\\2.0\\-0.5\\-15.9\end{array}\right]$&$\left[\begin{array}{c}1.8\\-3.5\\-0.4\\-0.5\end{array}\right]$\\\hline
\end{tabular}
\end{table}

For the $cn\bar{c}\bar{s}$ system, from Fig. \ref{fig-cncsbnbs}(a),
all the tetraquark states have rearrangement decay channels. Most of
the states have open-charm decay modes while the lowest $0^+$ and
$1^+$ not. Unlike the conventional mesons where the OZI rule works,
at present, we do not know whether the tetraquarks are broad or not
even if the state has only hidden-charm decay channels. From Table
\ref{cncsbnbs-Kij}, the highest $2^+$ and the second highest $0^+$
states should be relatively stable since the diquarks have
effectively attractive color-spin interactions while quark-antiquark
interactions are effectively repulsive. Probably the highest $1^+$
is also not very broad because of the weakly attractive $cn$
interaction. Further studies on decay widths can help to understand
the properties.

For the $bn\bar{b}\bar{s}$ system, from Fig. \ref{fig-cncsbnbs}(b),
one sees that the lowest $0^+$ and $1^+$ states do not have
open-bottom rearrangement decay modes while others have, a feature
similar to $cn\bar{c}\bar{s}$. From Table \ref{cncsbnbs-Kij}, in
these tetraquarks, possible relatively stable states are the highest
$2^+$, highest $1^+$, and the second highest $0^+$.

\begin{figure}[htpb]
\begin{tabular}{ccc}
\includegraphics[width=220pt]{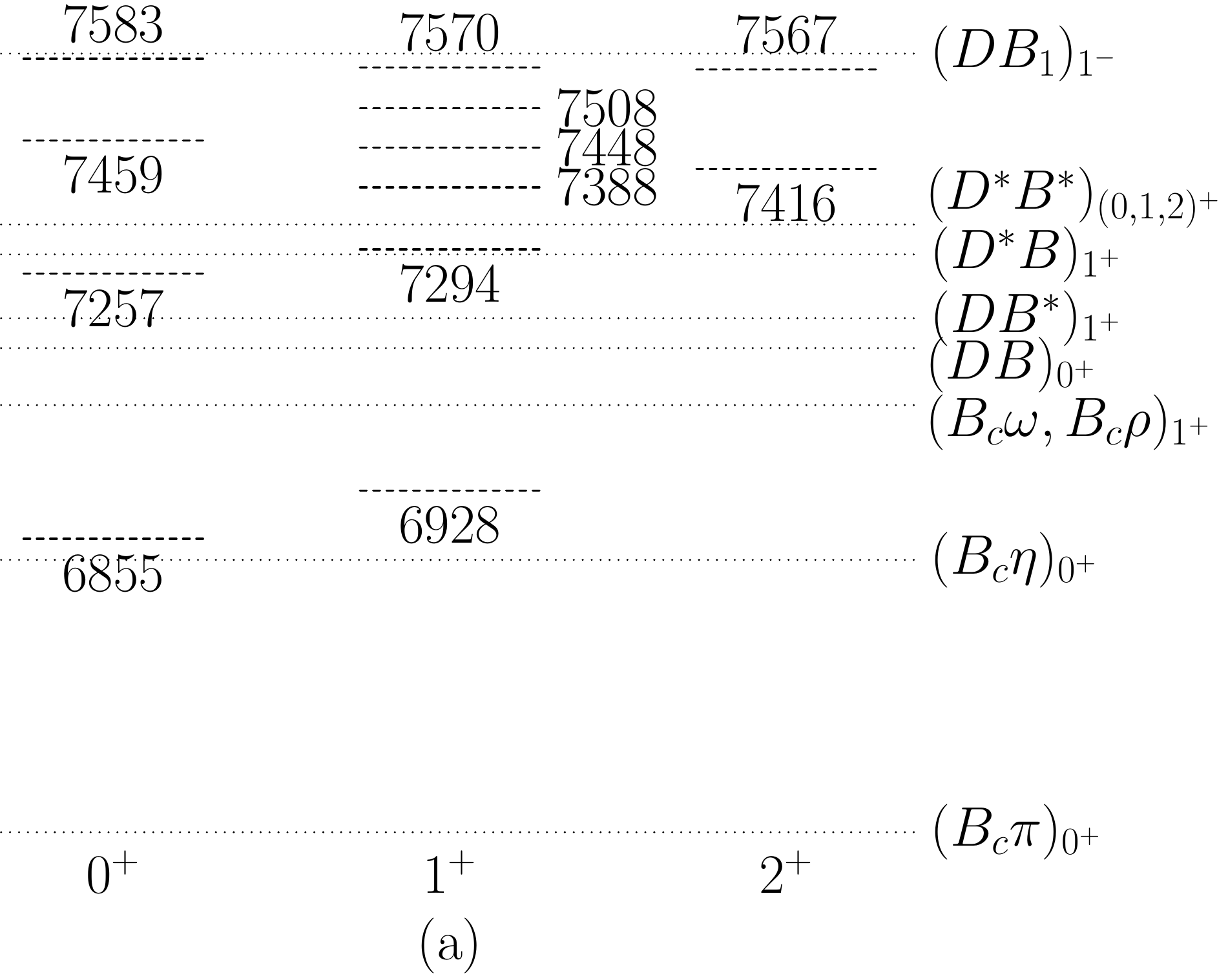}&\quad&
\includegraphics[width=220pt]{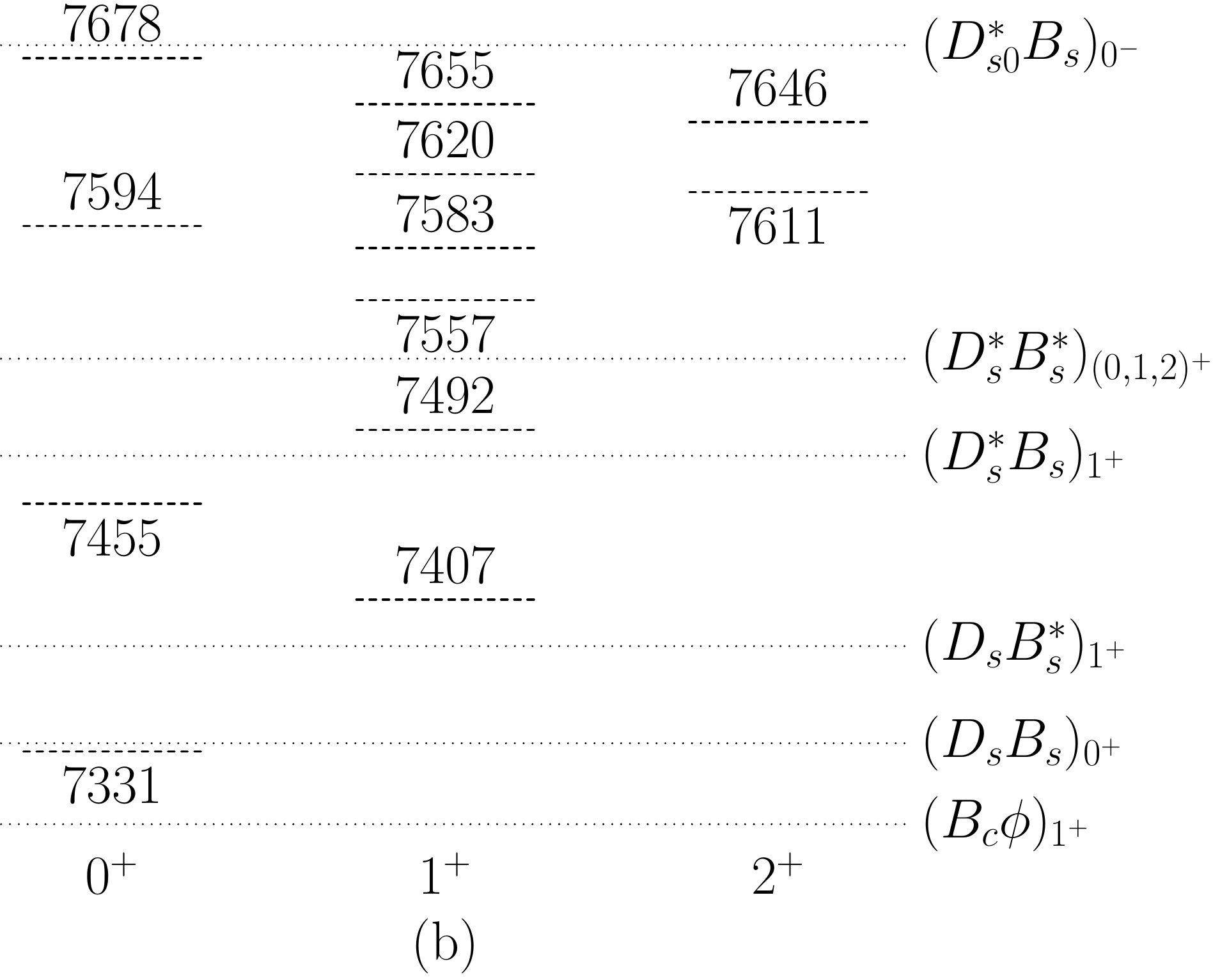}
\end{tabular}
\caption{Relative positions for $B_c$-like heavy tetraquarks (black
dashed lines): (a) $cn\bar{b}\bar{n}$ and (b) $cs\bar{b}\bar{s}$ and
various meson-meson thresholds (black dotted lines). The masses are
given in units of MeV. The subscripts of threshold symbols are
$J^{PC}$ in the $S$-wave case.}\label{fig-cnbncsbs}
\end{figure}

Now we move on to the $B_c$-like tetraquark states. With appropriate
substitutions of coupling parameters, one can obtain the eigenvalues
of the CMI matrices in Eqs. \eqref{matrixJ2}, \eqref{matrixJ0}, and
\eqref{matrixJ1}. Further, the tetraquark masses can be estimated in
various approaches mentioned above. We list these numerical results
in Table \ref{cnbncsbs}. The values in the last column are
determined with the help of the $X(4140)$ which is treated as a
$cs\bar{c}\bar{s}$ tetraquark. In this case, the mass formulas we
modify are
\begin{eqnarray}
M_{(cn\bar{b}\bar{n})}&=&m_b+m_c+2m_n+\langle
H_{CM}\rangle_{(cn\bar{b}\bar{n})}=(2m_c+2m_s)+(m_b-m_c)-2(m_s-m_n)+\langle
H_{CM}\rangle_{(cn\bar{b}\bar{n})},
\end{eqnarray}
and
\begin{eqnarray}
M_{(cs\bar{b}\bar{s})}&=&m_b+m_c+2m_s+\langle
H_{CM}\rangle_{(cs\bar{b}\bar{s})}=(2m_c+2m_s)+(m_b-m_c)+\langle
H_{CM}\rangle_{(cs\bar{b}\bar{s})}.
\end{eqnarray}
After the replacements we have used in previous systems are made,
one gets higher masses than those estimated with the $DB/D_sB_s$
threshold. We perform discussions with such masses. In Fig.
\ref{fig-cnbncsbs}(a) and \ref{fig-cnbncsbs}(b), we display the mass
spectra with dashed lines for $cn\bar{b}\bar{n}$ and
$cs\bar{b}\bar{s}$ tetraquarks, respectively. Relevant meson-meson
thresholds are also shown with dotted lines. The calculated
coefficients of effective color-spin interactions between quark
components, $K_{ij}$ in Eq. \eqref{eftKij}, are given in Table
\ref{cnbncsbs-Kij}.

\begin{table}[htbp]
\caption{Calculated CMI eigenvalues and estimated tetraquark masses
for the $cn\bar{b}\bar{n}$ and $cs\bar{b}\bar{s}$ systems in units
of MeV. The masses in the forth column (Upper limits) are obtained
with Eq. \eqref{mass} and those in the last three columns with
various reference states.}\label{cnbncsbs}
\begin{tabular}{c|cccccccccc}\hline
\multicolumn{7}{c}{$cn\bar{b}\bar{n}$ system} \\ \hline \hline
$J^{P}$&$\langle H_{CM}\rangle$&  Eigenvalues& Upper limits &
$B_c\pi$& $DB$ & X(4140)   \\ \hline
$2^{+}$&$\left(\begin{array}{cc}70.0&-68.7\\-68.7&132.6\end{array}\right)$&$\left(\begin{array}{c}176.8\\25.8\end{array}\right)$&$\left(\begin{array}{c}7678\\7527\end{array}\right)$&$\left(\begin{array}{c}7121\\6970\end{array}\right)$&$\left(\begin{array}{c}7462\\7311\end{array}\right)$&$\left(\begin{array}{c}7567\\7416\end{array}\right)$\\
$1^{+}$&$\left(\begin{array}{cccccc}-41.7&-58.6&41.3&68.7&87.6&-124.4\\-58.6&0.3&32.4&87.6&0.0&-118.5\\41.3&32.4&-28.5&-124.4&-118.5&0.0\\68.7&87.6&-124.4&-146.7&-146.6&103.2\\87.6&0.0&-118.5&-146.6&-0.1&81.0\\-124.4&-118.5&0.0&103.2&81.0&14.3\end{array}\right)$&$\left(\begin{array}{c}179.9\\118.4\\58.1\\-1.5\\-95.5\\-462.0\end{array}\right)$&$\left(\begin{array}{c}7681\\7619\\7559\\7499\\7405\\7039\end{array}\right)$&$\left(\begin{array}{c}7124\\7063\\7002\\6943\\6849\\6482\end{array}\right)$&$\left(\begin{array}{c}7465\\7403\\7343\\7283\\7189\\6823\end{array}\right)$&$\left(\begin{array}{c}7570\\7508\\7448\\7388\\7294\\6928\end{array}\right)$\\
$0^{+}$&$\left(\begin{array}{cccc}-97.6&137.5&-56.1&205.3\\137.5&-286.4&205.3&-140.3\\-56.1&205.3&-42.4&0.0\\205.3&-140.3&0.0&21.2\end{array}\right)$&$\left(\begin{array}{c}193.4\\69.0\\-132.8\\-534.8\end{array}\right)$&$\left(\begin{array}{c}7694\\7570\\7368\\6966\end{array}\right)$&$\left(\begin{array}{c}7138\\7013\\6811\\6409\end{array}\right)$&$\left(\begin{array}{c}7478\\7354\\7152\\6750\end{array}\right)$&$\left(\begin{array}{c}7583\\7459\\7257\\6855\end{array}\right)$\\
\hline\hline \multicolumn{7}{c}{$cs\bar{b}\bar{s}$ system} \\ \hline
$J^{P}$&$\langle H_{CM}\rangle$&  Eigenvalues& Upper limits & $B_c \phi$& $D_s B_s$  & X(4140)   \\ \hline
$2^{+}$&$\left(\begin{array}{cc}45.6&-13.7\\-13.7&68.5\end{array}\right)$&$\left(\begin{array}{c}74.9\\39.3\end{array}\right)$&$\left(\begin{array}{c}7933\\7897\end{array}\right)$&$\left(\begin{array}{c}7366\\7330\end{array}\right)$&$\left(\begin{array}{c}7554\\7518\end{array}\right)$&$\left(\begin{array}{c}7646\\7611\end{array}\right)$\\
$1^{+}$&$\left(\begin{array}{cccccc}-15.2&-21.9&5.3&13.7&11.3&-46.5\\-21.9&2.4&6.4&11.3&0.0&-64.6\\5.3&6.4&-32.8&-46.5&-64.6&0.0\\13.7&11.3&-46.5&-83.7&-54.8&13.3\\11.3&0.0&-64.6&-54.8&-1.2&16.1\\-46.5&-64.6&0.0&13.3&16.1&16.4\end{array}\right)$&$\left(\begin{array}{c}83.2\\48.7\\11.4\\-14.3\\-79.0\\-164.1\end{array}\right)$&$\left(\begin{array}{c}7941\\7907\\7869\\7844\\7779\\7694\end{array}\right)$&$\left(\begin{array}{c}7374\\7340\\7302\\7277\\7212\\7127\end{array}\right)$&$\left(\begin{array}{c}7562\\7528\\7491\\7465\\7400\\7315\end{array}\right)$&$\left(\begin{array}{c}7655\\7620\\7583\\7557\\7492\\7407\end{array}\right)$\\
$0^{+}$&$\left(\begin{array}{cccc}-45.7&27.3&-11.1&111.8\\27.3&-159.8&111.8&-27.9\\-11.1&111.8&-45.6&0.0\\111.8&-27.9&0.0&22.8\end{array}\right)$&$\left(\begin{array}{c}106.5\\23.0\\-116.9\\-240.9\end{array}\right)$&$\left(\begin{array}{c}7964\\7881\\7741\\7617\end{array}\right)$&$\left(\begin{array}{c}7398\\7314\\7174\\7050\end{array}\right)$&$\left(\begin{array}{c}7586\\7502\\7362\\7238\end{array}\right)$&$\left(\begin{array}{c}7678\\7594\\7455\\7331\end{array}\right)$\\
\hline
\end{tabular}
\end{table}

\begin{table}[htbp]
\caption{$K_{ij}$'s for $cn\bar{b}\bar{n}$ and $cs\bar{b}\bar{s}$
states. The order of states is the same as that in Table
\ref{cnbncsbs}.}\label{cnbncsbs-Kij}\tiny
\begin{tabular}{ccccccc}\hline
&\multicolumn{6}{c}{$cn\bar{b}\bar{n}$ system} \\ \hline \hline
$J^P$& $K_{cn}$ & $K_{b\bar{c}}$ & $K_{c\bar{n}}$ & $K_{b\bar{n}}$ & $K_{n\bar{n}}$ & $K_{bn}$ \\
$2^{+}$&$\left[\begin{array}{c}-0.2\\1.5\end{array}\right]$&$\left[\begin{array}{c}5.3\\-0.7\end{array}\right]$&$\left[\begin{array}{c}0.2\\4.5\end{array}\right]$&$\left[\begin{array}{c}0.2\\4.5\end{array}\right]$&$\left[\begin{array}{c}5.3\\-0.7\end{array}\right]$&$\left[\begin{array}{c}-0.2\\1.5\end{array}\right]$\\
$1^{+}$&$\left[\begin{array}{c}0.9\\1.0\\-3.4\\1.1\\-0.4\\-0.6\end{array}\right]$&$\left[\begin{array}{c}5.1\\-12.8\\-3.2\\1.1\\-0.1\\5.3\end{array}\right]$&$\left[\begin{array}{c}0.9\\1.6\\1.9\\4.0\\-12.5\\-0.6\end{array}\right]$&$\left[\begin{array}{c}-0.7\\2.1\\0.9\\-10.5\\3.5\\0.1\end{array}\right]$&$\left[\begin{array}{c}5.2\\4.7\\2.3\\-0.5\\-0.6\\-15.9\end{array}\right]$&$\left[\begin{array}{c}-0.9\\1.6\\-0.9\\-0.7\\-0.5\\0.1\end{array}\right]$\\
$0^{+}$&$\left[\begin{array}{c}2.3\\-4.3\\0.0\\-0.7\end{array}\right]$&$\left[\begin{array}{c}4.9\\2.1\\-0.6\\-15.8\end{array}\right]$&$\left[\begin{array}{c}2.1\\2.4\\-13.0\\-0.9\end{array}\right]$&$\left[\begin{array}{c}2.1\\2.4\\-13.0\\-0.9\end{array}\right]$&$\left[\begin{array}{c}4.9\\2.1\\-0.6\\-15.8\end{array}\right]$&$\left[\begin{array}{c}2.3\\-4.3\\0.0\\-0.7\end{array}\right]$\\
\hline
\end{tabular}
\begin{tabular}{ccccccc}\hline
&\multicolumn{6}{c}{$cs\bar{b}\bar{s}$ system} \\ \hline \hline
$J^{P}$ & $K_{cs}$ & $K_{b\bar{c}}$ & $K_{c\bar{s}}$ & $K_{b\bar{s}}$ & $K_{s\bar{s}}$ & $K_{bs}$ \\
$2^{+}$&$\left[\begin{array}{c}-0.6\\2.0\end{array}\right]$&$\left[\begin{array}{c}5.1\\-0.5\end{array}\right]$&$\left[\begin{array}{c}0.8\\3.9\end{array}\right]$&$\left[\begin{array}{c}0.8\\3.9\end{array}\right]$&$\left[\begin{array}{c}5.1\\-0.5\end{array}\right]$&$\left[\begin{array}{c}-0.6\\2.0\end{array}\right]$\\
$1^{+}$&$\left[\begin{array}{c}2.8\\-2.5\\2.6\\-4.0\\2.2\\-2.4\end{array}\right]$&$\left[\begin{array}{c}3.7\\-1.0\\-4.0\\-8.3\\0.2\\4.6\end{array}\right]$&$\left[\begin{array}{c}2.1\\3.7\\1.7\\2.6\\-10.7\\-4.0\end{array}\right]$&$\left[\begin{array}{c}-0.4\\4.2\\-4.3\\-8.5\\3.0\\1.2\end{array}\right]$&$\left[\begin{array}{c}4.6\\2.4\\1.2\\3.0\\-2.1\\-13.8\end{array}\right]$&$\left[\begin{array}{c}-2.7\\2.9\\-1.2\\0.8\\-1.9\\0.6\end{array}\right]$\\
$0^{+}$&$\left[\begin{array}{c}3.4\\-5.9\\2.3\\-2.4\end{array}\right]$&$\left[\begin{array}{c}4.1\\2.5\\-2.3\\-13.7\end{array}\right]$&$\left[\begin{array}{c}3.3\\2.4\\-10.9\\-4.2\end{array}\right]$&$\left[\begin{array}{c}3.3\\2.4\\-10.9\\-4.2\end{array}\right]$&$\left[\begin{array}{c}4.1\\2.5\\-2.3\\-13.7\end{array}\right]$&$\left[\begin{array}{c}3.4\\-5.9\\2.3\\-2.4\end{array}\right]$\\
\hline
\end{tabular}
\end{table}

Again one should note that the masses for $cn\bar{b}\bar{n}$ states
in the present model correspond to both the $I=1$ case and the $I=0$
case. From Fig. \ref{fig-cnbncsbs}, the lowest isoscalar $0^+$ state
seems to be narrow if it exists while its isovector partner may be
broader. If the mass difference $m_{B_c^*}-m_{B_c}=70$ MeV
\cite{Godfrey:1985xj} is used, the lowest $1^+$ tetraquark may decay
into $B_c^*\pi$ or $B_c^*\eta$. The isovector state seems to have a
broader width than its isoscalar partner. From Table
\ref{cnbncsbs-Kij}, the highest $1^+$ states and the second highest
$0^+$ states probably have relatively stable structures (for both
$I=1$ and $I=0$ cases).

For the $cs\bar{b}\bar{s}$ states, the lowest $0^+$ is around the threshold of
$D_sB_s$ and probably not broad, while the lowest $1^+$ may decay into $D_sB_s^*$,
$B_c\phi$, and $B_c^*\phi$. Other tetraquarks should have broader widths.
However, from Table \ref{cnbncsbs-Kij}, the highest $2^+$ and the second highest
$0^+$ probably have relatively stable structures.

In the hidden-charm (hidden-bottom) case, the minimal excitation
energy for the creation of a light quark-antiquark pair is around
370 (740) MeV while the orbital excitation energy is around 300
(400) MeV. In the present $B_c$ case, the excitation energy for a
light quark-antiquark pair is around 570 MeV, a value between the
hidden-charm and hidden-bottom cases. That for orbital excitation
should be less than 400 MeV, e.g. 370 MeV. Then the mass of
$B_{c0}(1P)$ is probably around 6.7 GeV and the mass of $B_{c2}(1P)$
is likely to be less than 6.8 GeV. From the QM calculations
\cite{Godfrey:1985xj}, we may also guess that the mass for the
$B_{c2}(1F)$ meson is probably in the range $7.2\sim7.3$ GeV. From
these numbers, it seems that only radially excited $B_c$ states with
$J^P=0^+$, $1^+$, and $2^+$ can fall into the mass region for the
$B_c$-like tetraquarks.

\subsection{The $cn\bar{b}\bar{s}$ and $cs\bar{b}\bar{n}$ systems }\label{sec3.4}

\begin{table}[!h]
\caption{Calculated CMI eigenvalues and estimated tetraquark masses
for the $cn\bar{b}\bar{s}$ and $cs\bar{b}\bar{n}$ systems in units
of MeV. The masses in the forth column (Upper limits) are obtained
with Eq. \eqref{mass} and those in the last three columns with
various reference states.}\label{cnbscsbn}
\begin{tabular}{c|cccccccccc}\hline
\multicolumn{7}{c}{$cn\bar{b}\bar{s}$ system} \\ \hline \hline
$J^{P}$ &$\langle H_{CM}\rangle$&  Eigenvalues& Upper limits & $B_c K$ & $D_s B$ & X(4140)   \\ 
$2^{+}$&$\left(\begin{array}{cc}54.9&-37.3\\-37.3&95.7\end{array}\right)$&$\left(\begin{array}{c}117.9\\32.8\end{array}\right)$&$\left(\begin{array}{c}7797\\7712\end{array}\right)$&$\left(\begin{array}{c}7238\\7153\end{array}\right)$&$\left(\begin{array}{c}7506\\7421\end{array}\right)$&$\left(\begin{array}{c}7598\\7513\end{array}\right)$\\
$1^{+}$&$\left(\begin{array}{cccccc}-27.2&-37.7&20.4&37.3&43.2&-80.0\\-37.7&1.1&17.6&43.2&0.0&-87.1\\20.4&17.6&-28.8&-80.0&-87.1&0.0\\37.3&43.2&-80.0&-109.6&-94.3&50.9\\43.2&0.0&-87.1&-94.3&-0.5&44.0\\-80.0&-87.1&0.0&50.9&44.0&14.4\end{array}\right)$&$\left(\begin{array}{c}122.5\\72.0\\29.2\\1.4\\-89.0\\-286.6\end{array}\right)$&$\left(\begin{array}{c}7802\\7751\\7709\\7681\\7590\\7393\end{array}\right)$&$\left(\begin{array}{c}7243\\7193\\7150\\7122\\7032\\6834\end{array}\right)$&$\left(\begin{array}{c}7511\\7460\\7418\\7390\\7299\\7102\end{array}\right)$&$\left(\begin{array}{c}7603\\7553\\7510\\7482\\7392\\7194\end{array}\right)$\\
$0^{+}$&$\left(\begin{array}{cccc}-68.3&74.7&-30.5&150.9\\74.7&-212.3&150.9&-76.2\\-30.5&150.9&-41.6&0.0\\150.9&-76.2&0.0&20.8\end{array}\right)$&$\left(\begin{array}{c}140.1\\45.0\\-125.8\\-360.7\end{array}\right)$&$\left(\begin{array}{c}7819\\7724\\7554\\7319\end{array}\right)$&$\left(\begin{array}{c}7261\\7166\\6995\\6760\end{array}\right)$&$\left(\begin{array}{c}7528\\7433\\7263\\7028\end{array}\right)$&$\left(\begin{array}{c}7621\\7526\\7355\\7120\end{array}\right)$\\
\hline\hline
\multicolumn{7}{c}{$cs\bar{b}\bar{n}$ system} \\ \hline
$J^{P}$&$\langle H_{CM}\rangle$ &Eigenvalues & Upper limits& $B_c\bar{K}$& $D B_s$  & X(4140)   \\
$2^{+}$&$\left(\begin{array}{cc}56.8&-36.8\\-36.8&95.6\end{array}\right)$&$\left(\begin{array}{c}117.8\\34.6\end{array}\right)$&$\left(\begin{array}{c}7797\\7714\end{array}\right)$&$\left(\begin{array}{c}7238\\7155\end{array}\right)$&$\left(\begin{array}{c}7493\\7410\end{array}\right)$&$\left(\begin{array}{c}7598\\7515\end{array}\right)$\\
$1^{+}$&$\left(\begin{array}{cccccc}-25.9&-37.3&20.7&36.8&44.0&-79.2\\-37.3&1.6&17.3&44.0&0.0&-87.7\\20.7&17.3&-32.5&-79.2&-87.7&0.0\\36.8&44.0&-79.2&-111.1&-93.3&51.9\\44.0&0.0&-87.7&-93.3&-0.8&43.3\\-79.2&-87.7&0.0&51.9&43.3&16.3\end{array}\right)$&$\left(\begin{array}{c}122.9\\72.9\\28.3\\-1.3\\-88.2\\-287.1\end{array}\right)$&$\left(\begin{array}{c}7802\\7752\\7708\\7678\\7591\\7392\end{array}\right)$&$\left(\begin{array}{c}7244\\7194\\7149\\7119\\7032\\6834\end{array}\right)$&$\left(\begin{array}{c}7499\\7449\\7404\\7374\\7288\\7089\end{array}\right)$&$\left(\begin{array}{c}7604\\7554\\7509\\7479\\7392\\7194\end{array}\right)$\\
$0^{+}$&$\left(\begin{array}{cccc}-67.2&73.5&-30.0&151.9\\73.5&-214.4&151.9&-75.1\\-30.0&151.9&-46.4&0.0\\151.9&-75.1&0.0&23.2\end{array}\right)$&$\left(\begin{array}{c}142.4\\42.4\\-127.8\\-361.8\end{array}\right)$&$\left(\begin{array}{c}7822\\7722\\7552\\7318\end{array}\right)$&$\left(\begin{array}{c}7263\\7163\\6993\\6759\end{array}\right)$&$\left(\begin{array}{c}7518\\7418\\7248\\7014\end{array}\right)$&$\left(\begin{array}{c}7623\\7523\\7353\\7119\end{array}\right)$\\
\hline
\end{tabular}
\end{table}

These states are composed of four different flavors, a similar
feature to the $X(5568)$. Some results can be found in Ref.
\cite{Cui:2006mp}. In fact, the isovector $B_c$-like systems also
contain quarks with four different flavors.

\begin{figure}[htpb]
\begin{tabular}{ccc}
\includegraphics[width=220pt]{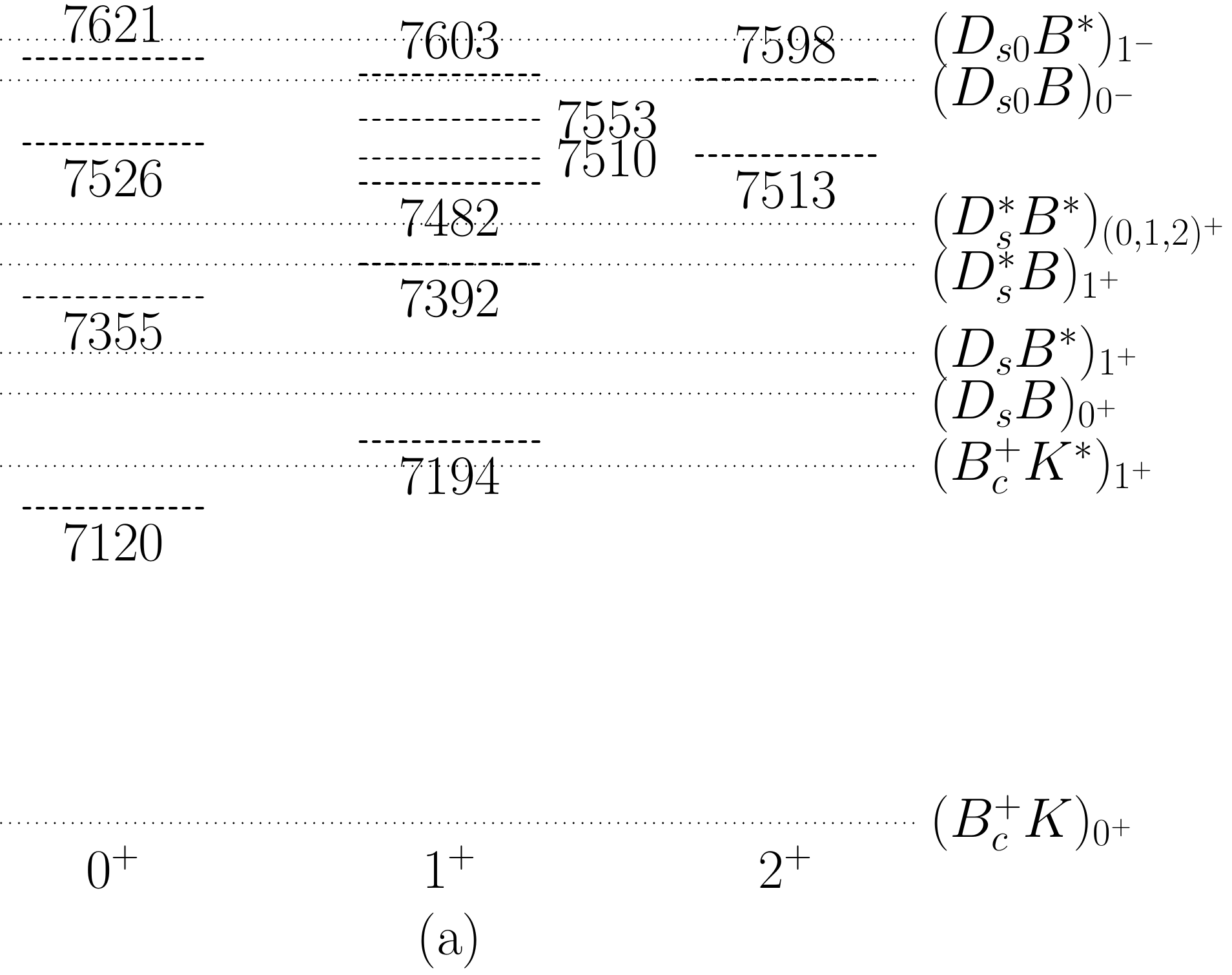}&\quad&
\includegraphics[width=220pt]{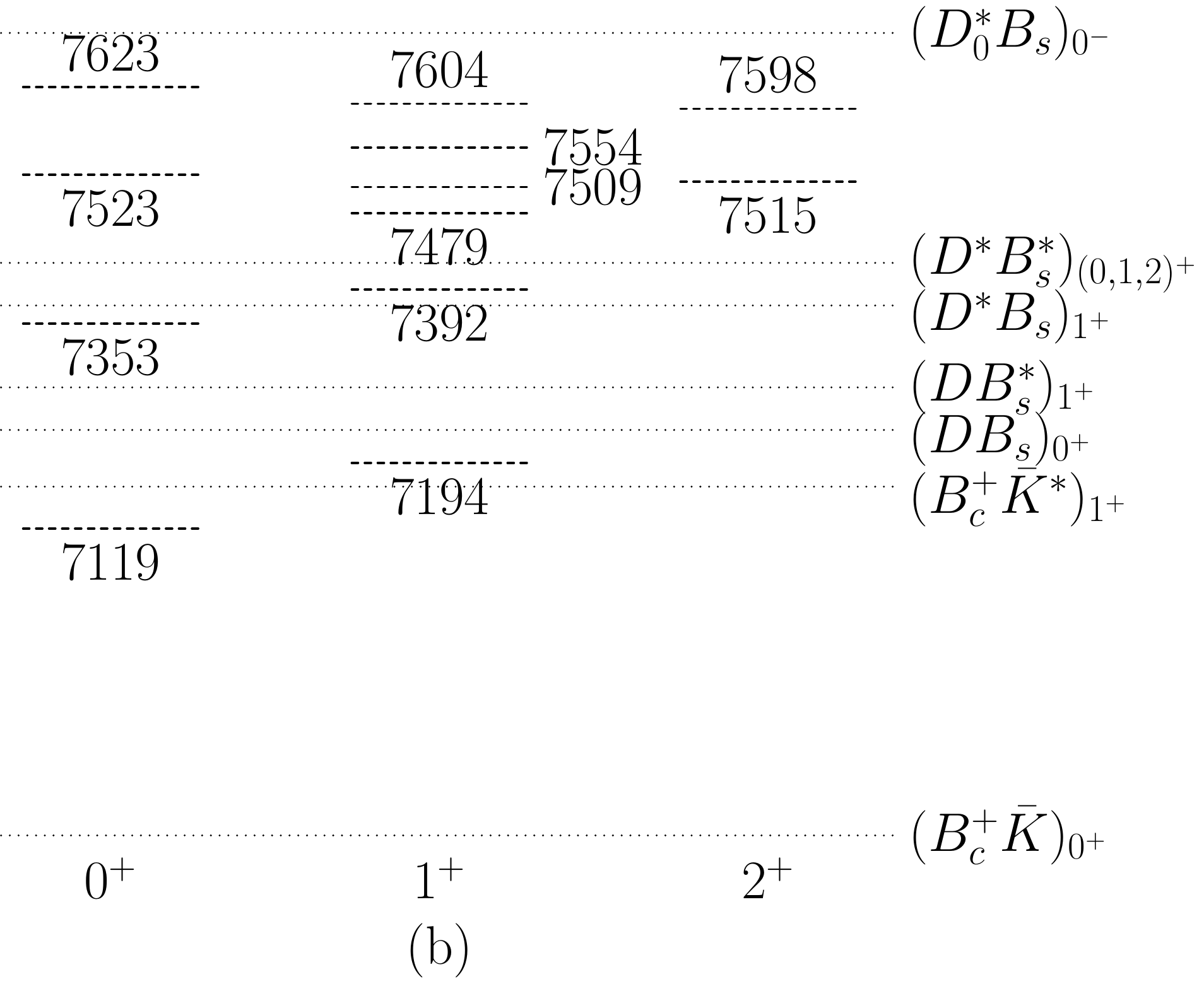}
\end{tabular}
\caption{Relative positions for $B_c$-like heavy tetraquarks (black
dashed lines): (a) $cn\bar{b}\bar{s}$ and (b) $cs\bar{b}\bar{n}$ and
various meson-meson thresholds (black dotted lines). The masses are
given in units of MeV. The subscripts of threshold symbols are
$J^{PC}$ in the $S$-wave case.}\label{fig-cnbscsbn}
\end{figure}

According to the expressions for $\langle H_{CM}\rangle$ in Eqs.
\eqref{matrixJ2}, \eqref{matrixJ0}, and \eqref{matrixJ1} and the
values of $C_{cn}$, $C_{b\bar{c}}$, $C_{c\bar{s}}$, $C_{b\bar{n}}$,
$C_{n\bar{s}}$, $C_{bs}$, $C_{cs}$, $C_{c\bar{n}}$, $C_{b\bar{s}}$,
and $C_{b\bar{n}}$ in Table \ref{parameter}, we obtain numerical
values and eigenvalues of the CMI matrices for $cn\bar{b}\bar{s}$
and $cs\bar{b}\bar{n}$ systems. These data together with the
estimated tetraquark masses in various approaches are shown in Table
\ref{cnbscsbn}. We have obtained masses with the help of the
$X(4140)$ in the last column by modifying
\begin{eqnarray}
M_{(cn\bar{b}\bar{s})}&=&m_b+m_c+m_n+ms+\langle
H_{CM}\rangle_{(cn\bar{b}\bar{s})}=(2m_c+2m_s)+(m_b-m_c)-(m_s-m_n)+\langle
H_{CM}\rangle_{(cn\bar{b}\bar{s})}
\end{eqnarray}
and
\begin{eqnarray}
M_{(cs\bar{b}\bar{n})}&=&m_b+m_c+m_n+m_s+\langle
H_{CM}\rangle_{(cs\bar{b}\bar{n})}=(2m_c+2m_s)+(m_b-m_c)-(m_s-m_n)+\langle
H_{CM}\rangle_{(cs\bar{b}\bar{n})}.
\end{eqnarray}
That is, the relevant formula is (to get more reasonable masses)
\begin{eqnarray}
&M=M_{X(4140)}-m_{B_s}-m_{D}+2m_{B}+\langle
H_{CM}\rangle_{B_s}+\langle H_{CM}\rangle_D-2\langle
H_{CM}\rangle_{B}-\langle H_{CM}\rangle_{X(4140)}+\langle
H_{CM}\rangle.
\end{eqnarray}
The mass differences between the $cn\bar{b}\bar{s}$ and
$cs\bar{b}\bar{n}$ states mainly come from the chromomagnetic
interactions. From Table \ref{cnbscsbn}, their differences are very
small. If we check the variables defined in Eq.
\eqref{definedvariables}, the differences in expressions are related
to $(C_{cn}-C_{cs})\pm(C_{bn}-C_{bs})$,
$(C_{c\bar{n}}-C_{c\bar{s}})\pm(C_{b\bar{n}}-C_{b\bar{s}})$, and
$-(C_{c\bar{n}}-C_{c\bar{s}})\pm(C_{b\bar{n}}-C_{b\bar{s}})$, i.e.
with the $SU(3)_f$ symmetry breaking when heavy quark is involved.
Numerically, Table \ref{parameter} tells us that their absolute
values are all less than 1 MeV and Table \ref{cnbscsbn} lets us know
that the resulting mass difference is at most 3 MeV. Because of the
existent heavy quark, the mass of a $c\bar{b}n\bar{s}$ is not
exactly the same as that of $c\bar{b}s\bar{n}$. We display the
relative positions for the tetraquark states and relevant thresholds
in Fig. \ref{fig-cnbscsbn}. One notes that mass differences
($10\sim20$ MeV) between $D_sB$ and $DB_s$, $D_sB^*$ and $DB_s^*$,
and so on also exist, but the properties of these two systems are
very similar. In Table \ref{cnbscsbn-Kij}, we present values of
$K_{ij}$'s for the present systems. The data for the two systems are
also close to each other. One may concentrate only on one system,
e.g. $cn\bar{b}\bar{s}$.

\begin{table}[!h]
\caption{$K_{ij}$'s for $cn\bar{b}\bar{s}$ and $cs\bar{b}\bar{n}$
states. The order of states is the same as that in Table
\ref{cnbscsbn}.}\label{cnbscsbn-Kij}\tiny
\begin{tabular}{ccccccc}\hline
&\multicolumn{6}{c}{$cn\bar{b}\bar{s}$ system} \\ \hline \hline
$J^{P}$  & $K_{cn}$ & $K_{b\bar{c}}$ & $K_{c\bar{s}}$ & $K_{b\bar{n}}$ & $K_{s\bar{n}}$ & $K_{bs}$ \\
$2^{+}$&$\left[\begin{array}{c}-0.3\\1.6\end{array}\right]$&$\left[\begin{array}{c}5.3\\-0.6\end{array}\right]$&$\left[\begin{array}{c}0.3\\4.3\end{array}\right]$&$\left[\begin{array}{c}0.3\\4.3\end{array}\right]$&$\left[\begin{array}{c}5.3\\-0.6\end{array}\right]$&$\left[\begin{array}{c}-0.3\\1.6\end{array}\right]$\\
$1^{+}$&$\left[\begin{array}{c}1.5\\-0.1\\-2.1\\-0.2\\0.5\\-1.0\end{array}\right]$&$\left[\begin{array}{c}4.8\\-6.9\\-7.5\\-0.1\\-0.1\\5.2\end{array}\right]$&$\left[\begin{array}{c}1.3\\3.0\\1.1\\3.9\\-12.7\\-1.2\end{array}\right]$&$\left[\begin{array}{c}-0.9\\3.8\\1.2\\-12.7\\3.7\\0.3\end{array}\right]$&$\left[\begin{array}{c}5.1\\3.5\\2.9\\0.1\\-0.6\\-15.6\end{array}\right]$&$\left[\begin{array}{c}-1.5\\2.3\\-1.6\\0.1\\-0.9\\0.2\end{array}\right]$\\
$0^{+}$&$\left[\begin{array}{c}2.9\\-5.2\\0.9\\-1.3\end{array}\right]$&$\left[\begin{array}{c}4.6\\2.3\\-0.8\\-15.5\end{array}\right]$&$\left[\begin{array}{c}2.7\\2.5\\-12.9\\-1.6\end{array}\right]$&$\left[\begin{array}{c}2.7\\2.5\\-12.9\\-1.6\end{array}\right]$&$\left[\begin{array}{c}4.6\\2.3\\-0.8\\-15.5\end{array}\right]$&$\left[\begin{array}{c}2.9\\-5.2\\0.9\\-1.3\end{array}\right]$\\
\hline
\end{tabular}
\begin{tabular}{ccccccc}\hline
&\multicolumn{6}{c}{$cs\bar{b}\bar{n}$ system} \\ \hline \hline
$J^{P}$& $K_{cs}$ & $K_{b\bar{c}}$ & $K_{c\bar{n}}$ & $K_{b\bar{s}}$ & $K_{s\bar{n}}$ & $K_{bn}$ \\
$2^{+}$&$\left[\begin{array}{c}-0.3\\1.6\end{array}\right]$&$\left[\begin{array}{c}5.3\\-0.6\end{array}\right]$&$\left[\begin{array}{c}0.3\\4.4\end{array}\right]$&$\left[\begin{array}{c}0.3\\4.4\end{array}\right]$&$\left[\begin{array}{c}5.3\\-0.6\end{array}\right]$&$\left[\begin{array}{c}-0.3\\1.6\end{array}\right]$\\
$1^{+}$&$\left[\begin{array}{c}1.7\\-0.1\\-1.8\\-0.5\\0.4\\-1.0\end{array}\right]$&$\left[\begin{array}{c}4.7\\-7.0\\-7.0\\-0.5\\-0.1\\5.2\end{array}\right]$&$\left[\begin{array}{c}1.4\\2.8\\1.2\\3.7\\-12.5\\-1.3\end{array}\right]$&$\left[\begin{array}{c}-0.8\\3.7\\1.3\\-12.7\\3.5\\0.3\end{array}\right]$&$\left[\begin{array}{c}5.1\\3.5\\2.7\\0.3\\-0.7\\-15.6\end{array}\right]$&$\left[\begin{array}{c}-1.6\\2.4\\-1.7\\0.3\\-0.9\\0.2\end{array}\right]$\\
$0^{+}$&$\left[\begin{array}{c}3.0\\-5.3\\0.9\\-1.3\end{array}\right]$&$\left[\begin{array}{c}4.6\\2.3\\-0.9\\-15.4\end{array}\right]$&$\left[\begin{array}{c}2.7\\2.4\\-12.7\\-1.8\end{array}\right]$&$\left[\begin{array}{c}2.7\\2.4\\-12.7\\-1.8\end{array}\right]$&$\left[\begin{array}{c}4.6\\2.3\\-0.9\\-15.4\end{array}\right]$&$\left[\begin{array}{c}3.0\\-5.3\\0.9\\-1.3\end{array}\right]$\\
\hline
\end{tabular}
\end{table}

If the mass difference $m_{B_c^*}-m_{B_c}$ is around 70 MeV
\cite{Godfrey:1985xj}, from Fig. \ref{fig-cnbscsbn}, all these
tetraquarks have rearrangement decay modes and probably are not
narrow states. From Table \ref{cnbscsbn-Kij}, the highest $2^+$ and
the second highest $0^+$ states have relatively stable structures.
The width of the lowest $0^+$ state is probably not very large since
it has only one rearrangement decay channel $B_cK$. To understand
whether such states exist or not and whether the adopted method is
reasonable or not, searching for them in possible decay channels is
a worthwhile work.

Now we move on to the problem about the nature of a state below the
$B_cK$ threshold. From Table \ref{cnbscsbn}, the lowest tetraquark
(6760 MeV) we can obtain is only $\sim10$ MeV lower than the $B_cK$
threshold. If a $B_cK$ molecule exists, the binding energy should be
small since the $B_cK$ interaction is weak (the small scattering
length $a_{\eta_c\pi}$ in Ref. \cite{Yokokawa:2006td} as a
reference). If experiments could observe a state below the threshold
with a large energy gap, a similar situation to the $X(5568)$, it
will be very difficult to understand its nature either in the
tetraquark picture or molecule picture. One gets a similar
conclusion for the $I=1$ $cn\bar{b}\bar{n}$ case.

\section{Discussions and summary}\label{sec4}

In this study, we systematically analyze the spectra of the possible
$Q_1q_2\bar{Q}_3\bar{q}_4$ ($Q=b,c$ and $q=n,s$ with $n=u,d$)
tetraquark states by using the CMI model. We use the
diquark-antiqiquark bases to construct the wave functions and
calculate the CMI matrices. After diagonalizing the matrices, the
eigenvalues irrelevant with base choice are obtained. Such values
determine the mass splittings between states with the same quark
content. Since the present model does not involve dynamics, one
cannot determine the absolute masses by solving the bound state
problem. To get numerical results of masses, we tried several
estimation approaches: (1) with Eq. \eqref{mass}, (2) with the
$(Q_1\bar{Q}_3)(q_2\bar{q}_4)$ type meson-meson threshold as a
reference, (3) with the $(Q_1\bar{q}_4)(q_2\bar{Q}_3)$ type
meson-meson threshold as a reference, and (4) with a postulated mass
scale relating to $X(4140)$ as a reference. In the first approach,
the obtained masses are always larger than those in other
approaches. For conventional hadrons, the obtained masses are
usually higher than the experimental measurements (see Table IV of
Ref. \cite{Zhou:2018pcv}). This means that the additional attraction
effects are actually needed in this approach and we may treat
tetraquark masses in this approach as theoretical upper limits. In
the second approach, the obtained masses are always smaller than
those in other approaches. Therefore, we may treat masses in this
approach as theoretical lower limits. In the third approach, the
obtained masses are moderate. Now we analyze the reason why the
masses in this approach are larger than those in the second
approach. In fact, when we estimating masses with the modified Eq.
\eqref{massref}, we are making the following replacements from Eq.
\eqref{mass},
\begin{eqnarray}
m_1+m_3&=&M^{Th.}_{(Q_1\bar{Q}_3)}-\langle H_{CM}\rangle_{(Q_1\bar{Q}_3)}\to M^{Ex.}_{(Q_1\bar{Q}_3)}-\langle H_{CM}\rangle_{(Q_1\bar{Q}_3)},\nonumber\\
m_2+m_4&=&M^{Th.}_{(q_2\bar{q}_4)}-\langle
H_{CM}\rangle_{(q_2\bar{q}_4)}\to M^{Ex.}_{(q_2\bar{q}_4)}-\langle
H_{CM}\rangle_{(q_2\bar{q}_4)}
\end{eqnarray}
in the second approach and
\begin{eqnarray}
m_1+m_4&=&M^{Th.}_{(Q_1\bar{q}_4)}-\langle H_{CM}\rangle_{(Q_1\bar{q}_4)}\to M^{Ex.}_{(Q_1\bar{q}_4)}-\langle H_{CM}\rangle_{(Q_1\bar{q}_4)},\nonumber\\
m_2+m_3&=&M^{Th.}_{(q_2\bar{Q}_3)}-\langle
H_{CM}\rangle_{(q_2\bar{Q}_3)}\to M^{Ex.}_{(q_2\bar{Q}_3)}-\langle
H_{CM}\rangle_{(q_2\bar{Q}_3)}
\end{eqnarray}
in the third approach. Here, $M^{Th.}_{(Q_1\bar{Q}_3)}$
($M^{Ex.}_{(Q_1\bar{Q}_3)}$) means the calculated (measured) mass
for the $(Q_1\bar{Q}_3)$ meson, etc. Then the compensated attraction
in the second approach is represented by
$(M^{Th.}_{(Q_1\bar{Q}_3)}-M^{Ex.}_{(Q_1\bar{Q}_3)})+(M^{Th.}_{(q_2\bar{q}_4)}-M^{Ex.}_{(q_2\bar{q}_4)})$
and that in the third approach is
$(M^{Th.}_{(Q_1\bar{q}_4)}-M^{Ex.}_{(Q_1\bar{q}_4)})+(M^{Th.}_{(q_2\bar{Q}_3)}-M^{Ex.}_{(q_2\bar{Q}_3)})$.
From Table IV of Ref. \cite{Zhou:2018pcv}, the former value is
usually larger than the latter value and their difference is the
mass difference for tetraquarks between the two approaches. For a
tetraquark state, its size should be larger than that of a
conventional meson, which means that the compensated attraction
should not be so strong like the value in the second approach.
Although the tetraquark masses in the third approach are larger, it
seems that they are still smaller than the realistic case. From Ref.
\cite{Park:2015nha}, additional kinetic energy may contribute and
lead to larger masses. The dynamical calculation in Ref.
\cite{Liu:2019zuc} also favors the argument that the estimated masses
with the reference thresholds are still small. Therefore, the values
obtained in the forth approach seem to be more realistic. The masses
are about $80\sim105$ MeV higher than those in the third approach.
Treating the $X(4140)$ as the lowest $1^{++}$ $cs\bar{c}\bar{s}$
tetraquark state, the problem of mass estimation becomes the problem
to determine quark mass differences in hadrons. Here, we assume that
the largest values in the fourth approach are closest to the
realistic tetraquark masses and perform discussions. Searching for
the various predicted states in this approach may help to test the
assumptions we adopt.

For the color-spin interactions between quark components in
multiquark states, the complicated structure mixing effects may
change their original properties, from attractive to repulsive or
from repulsive to attractive. For the tetraquark states studied in
this work, attractive diquarks while repulsive quark-antiquark pairs
are helpful for relatively stable states. To understand this
property for the quark interactions, we evaluated the measure,
$K_{ij}$ defined in Ref. \cite{Li:2018vhp}, for various states.

According to the numerical results, we performed the discussions in
the previous section. Our results on the exotic $XYZ$ states may be
summarized as follows:
\begin{itemize}
\item From the qualitative features of both mass and width, the newly observed $Z_c(4100)$ by LHCb seems to be a $0^{++}$ $cn\bar{c}\bar{n}$ tetraquark state.
\item From the consistency of mass and width with the $Z_c(4100)$, the $X(3860)$ observed by Belle may be another $0^{++}$ $cn\bar{c}\bar{n}$ tetraquark state.
\item The $Z_c(4200)$ observed by Belle is probably a $1^{+-}$ $cn\bar{c}\bar{n}$ tetraquark state.
\item The $Z_c(4250)$ can be a tetraquark but the quantum numbers cannot be assigned.
\item The $Z_c(3900)$, $X(3940)$, and $X(4160)$ are unlikely compact tetraquark states.
\item The $Z_c(4020)$ is unlikely a compact tetraquark, but seems to be the hidden-charm correspondence of the $Z_b(10650)$ with $J^{PC}=1^{+-}$.
\end{itemize}

Our predictions on possible tetraquarks can be found in Fig. 1 of
Ref. \cite{Wu:2016gas} and Figs. \ref{fig-cncn}, \ref{fig-bnbnbsbs},
\ref{fig-cncsbnbs}, \ref{fig-cnbncsbs}, and \ref{fig-cnbscsbn}.
There should exist relatively narrow tetraquarks, such as the lowest
$0^{++}$ $cs\bar{c}\bar{s}$ and $cn\bar{c}\bar{n}$. In particular,
from the signs of measure for effective quark interactions, we find
that for the highest $2^+$ and the second highest $0^+$ states, the
structures probably are more stable than other partner states with
the same $J^P$, because the quark-quark interactions in them are
effectively attractive while the quark-antiquark interactions are
effectively repulsive. For the case having $C$-parity, the highest
$1^{++}$ states also have such a property. The remaining states
having such a property are the highest $1^+$ $cn\bar{c}\bar{s}$ and
$bn\bar{b}\bar{s}$. The widths of these mentioned states are
probably not very broad although their masses are not low.

In the modified estimation method, the dominant uncertainties for
the tetraquark masses are partly remedied, but the uncertainties in
coupling parameters still exist, although the effects on mass
splittings may be small. It seems that one cannot solve this problem
without dynamical calculations. We wait for experimental
measurements to answer whether the extracted $C_{ij}$'s from the
conventional hadrons are actually applicable to multiquarks or not
and how large the induced uncertainties for mass splittings are.

In our study, we did not consider the generally mixed isoscalar
states of $Q_1n\bar{Q}_3\bar{n}$ and $Q_1s\bar{Q}_3\bar{s}$, but considered the
states similar to the $\omega$ and $\phi$ case. The mixing between
$Q_1n\bar{Q}_3\bar{n}$ and $Q_1s\bar{Q}_3\bar{s}$ surely affects the
spectrum. Once the predicted tetraquarks could be confirmed, one may
study this case if necessary.

Here we consider the compact tetraquark states. In the literature,
there are studies of various charmonium-, bottomonium-, and
$B_c$-like meson-meson molecules
\cite{Zhang:2009vs,Sun:2012sy,Albuquerque:2012rq}. Apparently, the
two configurations are difficult to distinguish just from the
quantum numbers. Since the distances between quark components in
these two configurations are different, the masses are not always
the same. To identify the inner structure to which an observed meson
belongs, the decay properties should be helpful.

An inconsistency about decay width might exist in our arguments. If
our argument about stability of states and the assignments for
tetraquark states are correct, the qualitative consistency between
the widths of $X(4140)$ and $X(4274)$ is satisfied. That for
$Z_c(4100)$, $X(3860)$, and $Z_c(4200)$ is also observed. However,
if we compare the widths of $cn\bar{c}\bar{n}$ tetraquarks and those
of $cs\bar{c}\bar{s}$ tetraquarks, the consistency seems a problem.
The widths of the former states are larger than 100 MeV while those
for the latter are at most tens of MeV. It is worthwhile to study
more on the decay widths of tetraquarks
\cite{Liu:2014eka,Ma:2015nmy} in future works in order to check or
confirm the assumptions used here.

To summarize, by studying the chromomagnetic interaction between
quark components, we calculated the mass splittings between the
$Q_1q_2\bar{Q}_3\bar{q}_4$ tetraquark states. With the assumption
that the $X(4140)$ is the lowest $1^{++}$ $cs\bar{c}\bar{s}$
tetraquark, we estimated all the $Q_1q_2\bar{Q}_3\bar{q}_4$ masses,
which can be tested in future experimental measurements. According
to the numerical results, we discussed possible assignments for
several exotic $XYZ$ mesons.

\section*{Acknowledgements}

J.W. thanks J.B. Cheng for checking some calculations. This project
is supported by Doctoral Research Fund of Shandong Jianzhu University (No. XNBS1851), National Natural Science Foundation of China (Grants Nos. 11775132, 11222547, 11175073, 11261130311, 11825503), and
973 program. X.L. is also the National Program for Support of
Top-notch Young Professionals.


\end{document}